\documentclass[manuscript, twocolumn]{aastex631}

\usepackage{color}

\let\tablewidth\undefined
\usepackage{tabularray}
\usepackage{graphicx}
\usepackage{amsmath}
\usepackage{gensymb}
\UseTblrLibrary{diagbox}

\newcommand\kepler{\textsl{Kepler}~}
\newcommand\tess{\textsl{TESS}~}

\newcommand{\esinw}{e\sin\omega}
\newcommand{\ecosw}{e\cos\omega}
\newcommand{\cosi}{\cos i}
\newcommand{\rsum}{(R_1+R_2)/a}
\newcommand{\rratio}{R_2/R_1}
\newcommand{\tratio}{T_2/T_1}

\received{2024 December 10}
\revised{2024 December 30}
\accepted{2024 December 30}
\published{2025 February 10}

\submitjournal{ApJS}

\begin{document}

\title{\vspace{0.2in} The Eclipsing Binaries via Artificial Intelligence. II. Need for Speed in PHOEBE Forward Models\vspace{0.5in}}

\shorttitle{The EBAI project. II.}
\shortauthors{Wrona \& Pr\v{s}a}

\correspondingauthor{Marcin Wrona}
\email{mwrona@villanova.edu}

\author[0000-0002-3051-274X]{Marcin Wrona}
\affiliation{Department of Astrophysics and Planetary Science, Villanova University, 800 East Lancaster Avenue, Villanova, PA 19085, USA}
\affiliation{Astronomical Observatory, University of Warsaw, Al. Ujazdowskie 4, 00-478 Warszawa, Poland}

\author[0000-0002-1913-0281]{Andrej Pr\v{s}a}
\affiliation{Department of Astrophysics and Planetary Science, Villanova University, 800 East Lancaster Avenue, Villanova, PA 19085, USA}

\begin{abstract}
	
In modern astronomy, the quantity of data collected has vastly exceeded the capacity for manual analysis, necessitating the use of advanced artificial intelligence (AI) techniques to assist scientists with the most labor-intensive tasks. AI can optimize simulation codes where computational bottlenecks arise from the time required to generate forward models. One such example is PHOEBE, a modeling code for eclipsing binaries (EBs), where simulating individual systems is feasible, but analyzing observables for extensive parameter combinations is highly time-consuming.

To address this, we present a fully connected feedforward artificial neural network (ANN) trained on a dataset of over one million synthetic light curves generated with PHOEBE. Optimization of the ANN architecture yielded a model with six hidden layers, each containing 512 nodes, providing an optimized balance between accuracy and computational complexity. Extensive testing enabled us to establish ANN's applicability limits and to quantify the systematic and statistical errors associated with using such networks for EB analysis. Our findings demonstrate the critical role of dilution effects in parameter estimation for EBs, and we outline methods to incorporate these effects in AI-based models.

This proposed ANN framework enables a speedup of over four orders of magnitude compared to traditional methods, with systematic errors not exceeding 1\%, and often as low as 0.01\%, across the entire parameter space.
	
\end{abstract}


\section{Introduction} \label{sec:intro}

Fundamental stellar properties are inferred predominantly from the study of eclipsing binary stars (EBs; \citealt{torres2010}). Their favorable orbital alignment with the line of sight, and consequent eclipses, make them ideal astrophysical laboratories: a simple geometry coupled with well-understood dynamical laws allows us to obtain fundamental parameters without a priori assumptions \citep{prsa2018}. In addition, an increasing number of EBs are found in triple and multiple systems \citep{conroy2014,orosz2015}, hosting circumbinary planets \citep{welsh2015}, and featuring mass transfer and apsidal motion \citep{hambleton2013}; these broaden the domains of study while retaining the same tractable modeling principles. In particular, we can probe stellar interiors by studying tidally induced oscillations and gravity-mode pulsations in detached binaries \citep{huber2015}; ubiquitous contact binaries are still a bit of a mystery in terms of their formation and evolution, but they seem to be associated with multiple systems \citep{tokovinin2014}; intriguing unique EBs, such as $\varepsilon$ Aur, VV Cep, V838 Mon, $\beta$ Cep and others, allow us to study accretion disks, colliding winds, quasiperiodic outbursts and reflective outer shells \citep{vogt2014}. Many of the phenomena being observed in hot Jupiters have their foundations in EB studies, e.g., the Rossiter–McLaughlin effect, tidal distortions of the host star, irradiation effects, Roche lobe overflow and wind outflows, gravity darkening, apsidal motion, third body dynamics, etc. \citep{barclay2012}. The study of EBs has been and continues to be vital for all of astrophysics.

Given the importance of EBs, it is probably not surprising why, in 1946, Henry Norris Russell described eclipses as the ``royal road'' to stellar astrophysics \citep{russell1948}. As royal as this road may be, it is also winding and bumpy. A modern EB modeling code will feature hundreds of parameters that describe the shapes, motions and interaction between the components. We use the model to synthesize high-accuracy, high-fidelity observables that can then be compared with actual observations \citep{wilson1971}. The most common EB observables are photometric light curves, high-resolution spectra, radial velocities, spectral line profiles, and astrometric positions \citep{prsa2018}.

This synthesis can be time-consuming: depending on parameter complexities and the number of timestamps in which observables should be synthesized, the computation can take anywhere from tens of seconds to tens of minutes or even longer. This should come as no surprise: general models are inherently numerical and need to do a lot. Stellar surfaces are distorted by tides, rotation, and obliquity and they need to be discretized; specific emergent intensity in the direction of the observer needs to be assigned to each surface element; this intensity is a function of the spatially variable temperature, surface gravity, chemical abundances, gravity/limb-darkening effects, and reflection; and the model needs to account for partial and/or total eclipses, ellipsoidal variability, and the circumstances will in general change from one point in orbit to another because of eccentric orbits \citep{prsa2018}. While simplified treatment can sometimes be warranted (for example, approximating stellar shapes with spheres for well-detached binaries), adopting such assumptions for the entire parameter space would inevitably lead to bias and, in turn, to systematically incorrect distributions of parameters. Rigor thus remains necessary, and computational time cost quite significant.

Yet this does not even begin to account for the time cost of the inverse problem: optimizing parameter values to get a match between synthesized observables and observations themselves. Similarly to many other problems with a large number of free parameters, the EB inverse problem is nonlinear and is inherently prone to parameter correlations and solution degeneracies \citep{prsa2005}. As a consequence, optimizers can only get us a solution, but not necessarily the solution. For that, we need samplers such as Markov Chain Monte Carlo (MCMC; \citealt{foremanmackey2017}) to provide heuristic parameter posteriors. This entails hundreds of thousands if not millions of forward-model runs, which puts a hard limit on the number of systems we can solve in a reasonable amount of time, and it necessitates the use of high-performance computing clusters (and an occasional bottle of wine) to get the job done.

Given these runtime constraints, the pertinent question is whether these forward-model calculations can be optimized, and whether the volume of these calculations can be reduced. Optimizing algorithms behind calculations cannot be done meaningfully—it is a complex problem that has already been heavily optimized over the last few decades. Calculation volume might be further reduced, but certainly not more than, say, 10\%–15\%. Thus, the truly impactful speedup must come from a fundamentally new approach. We present one such possibility in this paper.

\section{Artificial neural networks}

\begin{figure*}[t]
    \centering
    \includegraphics[width=\textwidth]{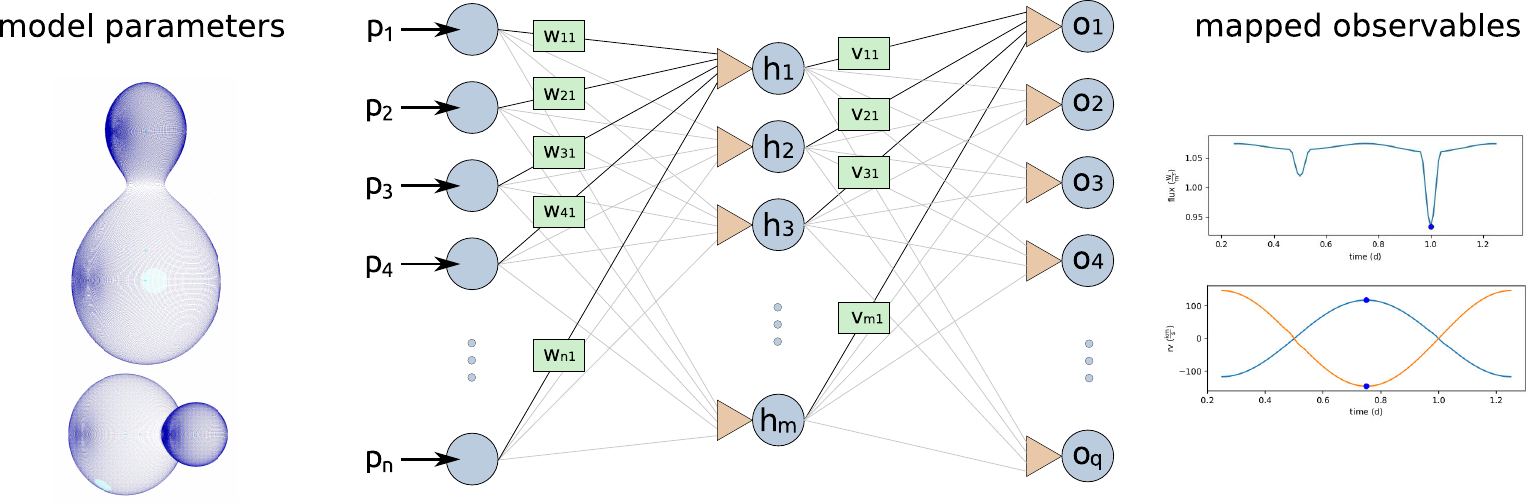} \\
    \caption{A schematic of the feedforward neural network. The IL (leftmost collection of nodes $p_{1}, \dots, p_{n}$) takes EB parameters as input; these parameters are then propagated to the hidden layer (middle collection of nodes $h_{1} \dots h_{m}$) and, in turn, to the OL (rightmost collection of nodes $o_{1} \dots o_{q}$). The mapping is uniquely defined by the set of weights $w_{11} \dots w_{nm}$, $v_{11} \dots v_{mq}$, and biases $b_{1}\dots b_{m}$, $c_{1}\dots c_{q}$ added after summation of weighted connections. These weights and biases are determined by backpropagation from the known combinations of model parameters and synthesized observables.}
    \label{fig:bpn}
\end{figure*}

Artificial neural networks (ANNs) are computational constructs that stem from the field of artificial intelligence (AI; \citealt{freeman1991}). An ANN is a stack of interconnected layers, visualized in Figure~\ref{fig:bpn}. Each layer is an array of processing elements called nodes, depicted in blue. Each node holds a single scalar value. The signal is propagated between layers by weighted connections, depicted in green; each receiving node takes a sum of all previous layer node values, each multiplied by the corresponding connection weight, plus an additional bias term: $h_k = \sum_i w_{ik} p_i + b_k$ and, similarly, $o_p = \sum_j v_{jp} h_j + c_p$. Finally, node values are typically passed through a nonlinear activation function (AF), depicted in amber. A typical choice for the AF is a $\texttt{sigmoid}$ function \citep{freeman1991}. Thus, the ANN performs a nonlinear mapping of input-layer (IL) data to the output-layer (OL) observables. In this work we focus on a feedforward neural network, which is a type of ANN where connections do not form cycles.

To demonstrate the ANN's principal use, consider a natural law that we can probe but cannot easily formulate. Thus, we can vary input parameters and measure the resulting quantities, but the actual underlying mechanism is unknown to us. Each measurement (a set of input parameters and the corresponding observed outcome) informs us of that underlying mechanism even if we do not know yet what it is. This is how we teach the ANN to ``compute'' the observables: a set of known parameters is fed into the IL, the forward mapping is determined by connection weights, and the predicted observables are harvested on the OL. We then compare these predictions with the actually measured observables and use the residuals to train the network: we adjust connection weights and biases so that the network reproduces the observed outcomes for all actual measurements as accurately as possible. In our particular case, the ``natural law'' is quite well understood: it is the sophisticated EB forward model. The main idea presented here is to replace expensive computations by using a trained ANN instead.

Say we knew the connection weights and biases that map EB parameters to synthetic observables to a high fidelity. In that case, going from parameters to observables is a matter of calculating $\sim10^3$ linear combinations of parameters on the IL, another $\sim10^5$ linear combinations on the hidden layers, evaluating a $\texttt{sigmoid}$ function $\sim10^3$ times, and obtaining the result. Thus, we need to evaluate a few hundred thousand time-inexpensive operations, which amounts to submilliseconds on any modern single processor. Compare fractions of milliseconds to seconds or minutes of actual forward-model computation time, and the motivation becomes crystal clear. Assuming the attained prediction fidelity, we will have gained $\sim5$ orders of magnitude in speed increase.

In the past, the ANNs were used in the exactly opposite way: observables were passed as input and the estimated parameters were harvested as output \citep{prsa2008}. The associated errors in parameter estimates were below 10\% for $\sim$90\% of the sample, but that was the ceiling in precision. In addition, such networks required data preprocessing—i.e., an analytical function that represents the data—which made the pipeline fragile \citep{slawson2011}. The ANN proposed here can be subject to the unchanged reverse problem infrastructure, including optimizing, heuristic sampling, etc., but with an $\sim$1 millionfold speedup. The price to pay is that we lock ourselves into the subsection of the parameter space used to train the network, but we demonstrate below that this is not as constraining as it might initially seem.

\section{Parameterizing the network}

In the absence of any auxiliary data or assumptions, EB light curves by themselves do not carry any information on the absolute scale of the system: we generally cannot infer masses, radii, luminosities, semimajor axes, individual component temperatures, stellar luminosities, or any kinematic properties of the binary system \citep{prsa2008}. We can, however, infer relative parameters, where quantities are expressed in the scale units of another quantity, such as the semimajor axis, or total luminosity of the system. Table \ref{tab:params} lists the most common EB parameters and the observables needed for parameter determination. Single light curves from, say, \kepler \citep{borucki2010} or \tess \citep{ricker2015} allow us to determine the sum of fractional radii ($R_1/a + R_2/a$), eccentricity and argument of periastron ($e$ and $\omega$), temperature ratio ($T_2/T_1$), and orbital inclination ($i$). In the case of total eclipses, we can also infer the ratio of radii ($R_2/R_1$); in the case of contact binaries, we can constrain the photometric mass ratio ($q = M_2/M_1$); for third light ($l_3$) we need robust detrending and nonvariable sources (even though there is some promise for detrending variable sources as well; \citealt{prsa2019}); and for temporally variable elements ($\dot P, \dot \omega, \dot i, \dot e, \dots$) we need dense and extended temporal coverage and smooth variations.

\begin{table*}[t!]
	\centering
	\caption{EB parameters that can be inferred from a specific combination of observables. ``Cal.~LC'' stands for flux-calibrated light curve, ``SB1'' stands for a single-lined spectroscopic binary, and ``SB2'' for the double-lined spectroscopic binary. The (\checkmark) symbol corresponds to quantities that can be inferred from some but not all light curves.}
	\scriptsize
	\begin{tabular}{|l|c|c|c|c|c|c|}
		\hline
		Parameter & Cal.~LC & Cal.~LC+distance & $2+$ Cal. LCs & $2+$ Cal. LCs+dist. & $2+$ Cal. LCs+SB1 & $2+$ Cal. LCs+SB2 \\
		\hline
		Conjunction time ($t_0$)                         & \checkmark & \checkmark & \checkmark & \checkmark & \checkmark & \checkmark \\
		Orbital period ($P_0$)                           & \checkmark & \checkmark & \checkmark & \checkmark & \checkmark & \checkmark \\
		Inclination ($i$)                                & \checkmark & \checkmark & \checkmark & \checkmark & \checkmark & \checkmark \\
		Tangential eccentricity ($e \cos \omega$)        & \checkmark & \checkmark & \checkmark & \checkmark & \checkmark & \checkmark \\
		Radial eccentricity ($e \sin \omega$)            & \checkmark & \checkmark & \checkmark & \checkmark & \checkmark & \checkmark \\
		Sum of fractional radii ($r_1+r_2$)              & \checkmark & \checkmark & \checkmark & \checkmark & \checkmark & \checkmark \\
		Temperature ratio ($T_2/T_1$)                    & \checkmark & \checkmark & \checkmark & \checkmark & \checkmark & \checkmark \\
		Radius ratio ($R_2/R_1$)                         & (\checkmark) & (\checkmark) & (\checkmark) & (\checkmark) & (\checkmark) & \checkmark \\
		Mass ratio ($q = M_2/M_1$)                       & (\checkmark) & (\checkmark) & (\checkmark) & (\checkmark) & (\checkmark) & \checkmark \\
		Third light ($l_3$)                              & (\checkmark) & (\checkmark) & (\checkmark) & (\checkmark) & (\checkmark) & (\checkmark) \\
		Variable elements ($\dot P, \dot \omega, \dots$) & (\checkmark) & (\checkmark) & (\checkmark) & (\checkmark) & (\checkmark) & (\checkmark) \\
		Sum of abs.~luminosities ($L_1 + L_2$)           &              & \checkmark   &              & \checkmark   &              & \checkmark   \\
		Absolute luminosities ($L_1$, $L_2$)             &              & \checkmark   &              & \checkmark   &              & \checkmark   \\
		Absolute temperatures ($T_1$, $T_2$)             &              &              & \checkmark   & \checkmark   & \checkmark   & \checkmark   \\
		Barycentric velocity ($v_\gamma$)                &              &              &              &              & \checkmark   & \checkmark   \\
		Absolute masses ($M_1$, $M_2$)                   &              &              &              &              &              & \checkmark   \\
		Absolute radii ($R_1$, $R_2$)                    &              &              &              & \checkmark   &              & \checkmark   \\
		Semi-major axis ($a$)                            &              &              &              & \checkmark   &              & \checkmark   \\
		\hline
	\end{tabular}
	\normalsize
	\label{tab:params}
\end{table*}

The absolute scale of the system is traditionally obtained by spectroscopic follow-up and extracting radial velocity shifts (RVs) for both components. RVs allow us to determine the semimajor axis, mass ratio, and barycentric velocity, which in turn ``absolutizes'' the model \citep{wilson1971}. Yet spectroscopic campaigns are time-intensive because we need a reasonable phase coverage to derive absolute parameters to a reasonable accuracy.

Another prospective data resource is Gaia \citep{gaia2021}. Gaia is an all-sky survey satellite that combines astrometry, photometry, spectrophotometry, and RV spectroscopy \citep{gaia2021}. While the onboard radial velocity instrument will provide a direct path to absolute parameters, the time-resolved RV measurements are not yet publicly available; they are expected to accompany Data Release 4 (DR4) in 2025 \citep{gaiadr}. We can, however, resort to the known distances from Gaia's parallax measurements \citep{bailerjones2018}. The reported mean uncertainty in parallax measurements is $\sim$40\,$\mu$as \citep{luri2018}, which corresponds to $\sim$4\% at a 1\,kpc distance. At the same time, both \kepler and \tess data are given in $\mathrm{e}^-/\mathrm{s}$, which (in conjunction with extinction maps) can be converted to absolute fluxes using dedicated tools \citep{ridden2021}, with the reported precision of $\sim$1\%. Thus, we can obtain absolute stellar luminosities to the precision of $\sim$6\%.

\section{Training the neural network} \label{sec:training}

To adequately train the ANN, we require a large precomputed dataset of observables from a set of model parameters. The computation of the dataset does take substantial computing time, but we only have to do this once. We can then use this dataset to train the ANN. This is another step that takes a long time, but it too needs to be done only once. From then onward, we can have the ANN ``compute'' forward models for millions of parameter combinations in a matter of seconds. Thus, once trained, the ANN holds the promise of delivering a millionfold speedup.

ANNs are particularly good at interpolating and particularly bad at extrapolating \citep{freeman1991}. In other words, the networks are expected to perform quite well on input parameters spanned by the training set, while their performance degrades rapidly off the grid. In addition, the ANN will not perform optimally if the density of the training sample is substantially different from that of the unseen sample: if we train the network on parameter combinations where only, say, 10\% exist in nature and the 90\% are bogus, the network will be poorly trained. That is why the training set needs to satisfy the following two criteria:

\begin{description}
	\item[The training set needs to be complete] The network is not able to synthesize observables it has not seen before; thus, the ranges of the parameters used for training the network represent a hard definition boundary for the unknown light curve set. If, during optimization or sampling, the values of the parameters exceed these boundaries, the solution is invalidated.
	
	\item[The training set needs to be representative] The combinations of parameters need to approximately match those that occur in nature. The emphasis on ``approximately'' is meant to imply that the similarity between the training set and the unknown set optimizes the performance rather than enables it: a poorly trained ANN can still perform reasonably well within its own completeness limits.
\end{description}

In order to assemble a robust training set, one should adopt the best parameter distribution estimates known to date \citep[e.g.,][]{raghavan2010,duchene2013,moe2017}, and use those distributions to synthesize a sample of binary stars \citep{arenou2011}, and then derive the distributions of EB parameters. These are then compared to the observed distributions from the data, and the starting distribution estimates are modified to minimize the discrepancies \citep{wells2017,wells2021}. The training set will follow those distributions, with a reasonable margin to ascertain completeness. As the network is trained and the resulting parameter values are validated, the training set will be revised in order to achieve completeness and ensure proper representation.

\subsection{The size of the training set}

The performance of the neural network will clearly depend on the properties of the training set. The size of the training set is inherently linked to the model it represents; in general, it will depend on the model complexity (number of parameters), nonlinearity (functional parameter relationships), and model uniqueness (parameter degeneracy). A recent review analyzed the literature to find that the vast majority of studies perform post hoc sample analysis and only a few studies attempted to estimate the size pre hoc \citep{balki2019}; even in those few cases, the analytical estimate proved to be of limited practical value. A typical rule of thumb to determine the sample size is 10 per degree of freedom \citep{sudipto2022}; thus, for six parameters, the estimated sample size would be $\sim$1 million. Yet our particular case has a distinct benefit over datasets where the underlying model is unknown: it unequivocally allows post festum analysis. Say we start with a sample of one million light curves. We can then train the network with this sample, use it to obtain a solution for the observed light curves, and compare the values of the parameters used for training to those derived from the analysis. Thus, we can measure the degree of overlap and, since the model is known and fully determined, we can modify and extend the training set accordingly. This is an iterative process that allows us to ascertain both completeness and appropriate representation density.

\subsection{PHOEBE models generator}
\label{sec:PHOEBE models generator}
To generate the models for the training sets, we utilize the most recent version of PHOEBE\footnote{At the time of this writing, PHOEBE 2.4.16.}. For our analysis, the parameters describing a binary system were grouped into two categories. The first set contains parameters that have the greatest impact on the shape of the light curve and are used as input parameters for the ANN. We decided to use the same set of parameters as presented in \cite{prsa2008} with one minor modification, as well as including an additional parameter. This set is presented in Table~\ref{table:parameters}. We will refer to this group as the main set of parameters. In Section 2.2 of \cite{prsa2008}, the authors explain how each of these parameters influences the shape of the light curve. Our only modification to the original set is the use of $\cos i$ instead of $\sin i$ due to the greater sensitivity of the cosine function for angles close to a right angle. The aforementioned additional parameter is the radius ratio $R_2/R_1$. This parameter is particularly significant in the case of total eclipses, which present themselves in the light curve as ``flat'' minima. The duration of such a total eclipse depends on $R_2/R_1$. This parameter also affects the amplitude of brightness variations, although this is a second-order effect. With the accuracy of parameter determinations provided by ANNs, it is possible to capture this effect.

\begin{table}[h]
	\centering
	\caption{Main parameter set with brief descriptions.}
	\begin{tabular}{ll}
		\hline
		\textbf{Parameter} & \textbf{Description} \\
		\hline
		$e\sin\omega$ & Tangential eccentricity \\
		$e\cos\omega$ & Radial eccentricity \\
		$\cos i$ & Cosine of inclination \\
		$(R_1+R_2)/a$ & Sum of relative radii \\
		$R_2/R_1$ & Radius ratio \\
		$T_2/T_1$ & Effective temperature ratio \\
		\hline
	\end{tabular}
	\label{table:parameters}
\end{table}

\begin{figure*}[t]
	\centering
	\includegraphics[width=0.8\linewidth]{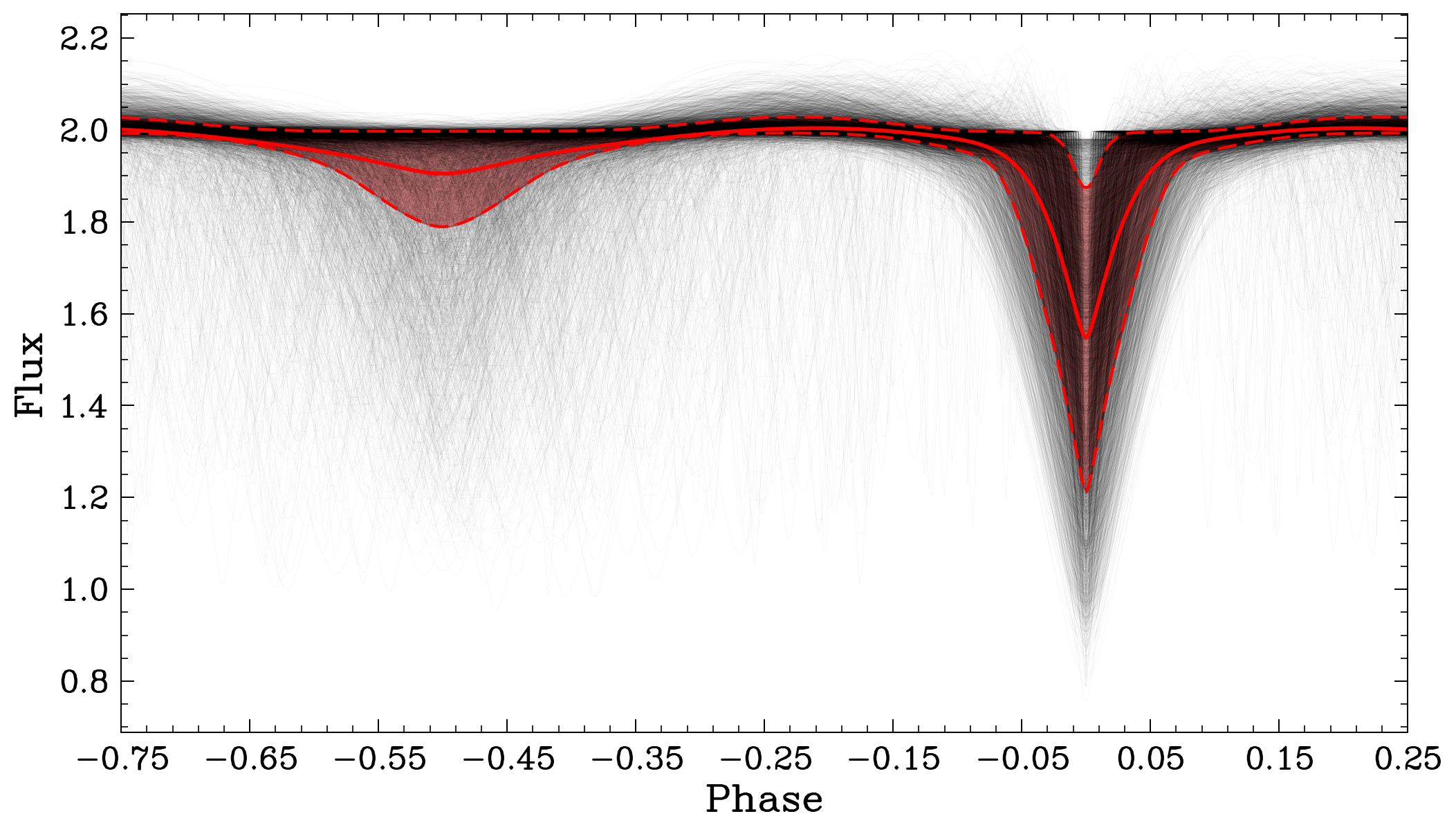}
	\caption{Overlay of light curves generated for a test sample. For each phase, the median and 18th–84th percentile range of fluxes were computed. The red solid line represents the median light curve, while the shaded region between the red dashed lines indicates the percentile range. Each thin black line represents an individual light curve. Although only 5000 out of the 1,250,000 generated light curves are shown here, the statistics were computed for the entire sample. See Section~\ref{sec:training_set_generation} for more details.
	}
	\label{fig:pmglcs}
\end{figure*}

The second set includes both orbital and physical parameters that describe the individual components and the system as a whole. These include the orbital period ($P$), the masses of both components ($M_1$ and $M_2$), the mass ratio ($q$), the logarithmic surface gravity of both components ($\log g_1$ and $\log g_2$), the primary radius ($R_1$), and its effective temperature ($T_1$). The code also accounts for atmospheric parameters and effects such as gravity darkening, irradiation/reflection, and limb darkening. Forward models were generated for a single passband -- TESS T \citep{ricker2015}. Given that some parameters are interdependent, specific parameters were fixed, with others determined through dependencies.

After selecting the parameters to fix and the ones to vary, the program randomly drew the main parameters from preestablished distributions. Since not all parameter combinations result in EBs, a filtering process was applied to ensure the physical validity of the system and the presence of eclipses in the light curve. Once a forward model was generated, both the set of main parameters and the corresponding light curves were saved to a data file, which was subsequently used to train the ANN.

We chose to parameterize the light curve as a 1D list of fluxes. Since all light curves were of equal length and evenly spaced in phase over a single orbital period, with the time of superior conjunction set to zero, the position in the list directly corresponds to the orbital phase. To limit the occurrence of eclipses at the boundaries of the chosen phase range, we selected a phase range from $-0.75$ to $0.25$. Within this range, the eclipses for most orbital parameter combinations are well separated from the edges, thereby improving the performance of the ANN. Figure~\ref{fig:pmglcs} shows a sample of the generated light curves plotted together. The plot also includes the median system and the 18th--84th percentile range of fluxes at each phase point. The median light curve corresponds to a system with relatively low eccentricity, featuring well-defined eclipses, a significant amplitude ratio, and noticeable ellipsoidal variability.

\subsection{Structure of the artificial neural network}
\label{sec:structure_of_DNN}
There is no universal rule for determining the optimal architecture of an ANN for a specific task. The number of possible configurations grows rapidly with the number of hidden layers (HLs), the number of nodes in those layers, and other training parameters such as batch size, number of epochs, activation functions (AFs), and the choice of the optimization algorithm. To define, train, and utilize a deep, fully connected ANN, we used \texttt{Keras} \citep{chollet2015keras}, a deep learning API written in Python, built on top of the machine learning platform \texttt{TensorFlow} \citep{tensorflow2015-whitepaper}.

The structure of the ANN required for our study is as follows. The input layer (IL) consists of six main orbital and physical parameters of the eclipsing system, as listed at the beginning of Section~\ref{sec:PHOEBE models generator}. Each node in the IL is fully connected to the nodes in the first HL, which, in turn, are fully connected to the nodes in the subsequent HLs, ultimately leading to the output layer (OL) that contains the predicted fluxes. Initially, we adopted a trial-and-error approach to determine a suitable ANN architecture. To optimize network architecture, we randomly sampled the parameter space using the \texttt{RandomizedSearchCV} function from the \texttt{scikit-learn} library \citep{scikit-learn}. \texttt{RandomizedSearchCV} (hereafter RSCV) selects a random subset of the parameter grid for evaluation. The \texttt{CV} part refers to cross-validation, a machine learning technique used to evaluate how well a model generalizes to new data. By default, it employs $K$-fold cross-validation, which involves splitting the data into $K$ equally sized folds. The algorithm trains the model $K$ times, each time using a different fold as the validation set and the remaining data as the training set. This process allows the algorithm to assess the model on multiple validation sets, providing a more robust estimate of its performance.

To evaluate the performance of our ANN architecture, we employed three different metrics:

\begin{itemize}
	\item \textbf{Mean Squared Error (MSE)}: This metric is sensitive to significant outliers or large prediction deviations.
	\begin{equation*}
		\mathrm{MSE} = \frac{1}{n} \sum_{i=1}^n (y_i - \hat{y}_i)^2
	\end{equation*}
	Here, \( n \) denotes the number of observations in the dataset, \( y_i \) is the actual value of the \( i \)-th flux point, and \( \hat{y}_i \) is the predicted value for the \( i \)-th flux point.
	
	\item \textbf{Median Absolute Error (MedAE)}: This metric is robust to outliers, providing an indication of the model's typical performance, even when a few predictions have large errors.
	\begin{equation*}
		\mathrm{MedAE} = \mathrm{median}(|y_1 - \hat{y}_1|, \dots, |y_n - \hat{y}_n|)
	\end{equation*}
	
	\item \textbf{Maximal Residual Sum of Squares (MaxRSS)}: The metric computes RSS for each predicted model and returns the maximum RSS value across all predicted light curves. This metric highlights the worst-case scenario, ensuring that even the largest prediction errors are accounted for.
	\begin{equation*}
		\mathrm{MaxRSS} = \max\left(\sum_{j=1}^m (y_j - \hat{y}_j)^2\right)
	\end{equation*}
	Here, the summation is calculated for each light curve, and \( m \) represents the number of flux points within a single light curve.
\end{itemize}

\begin{figure}[h]
	\centering
	\includegraphics[width=1.0\linewidth]{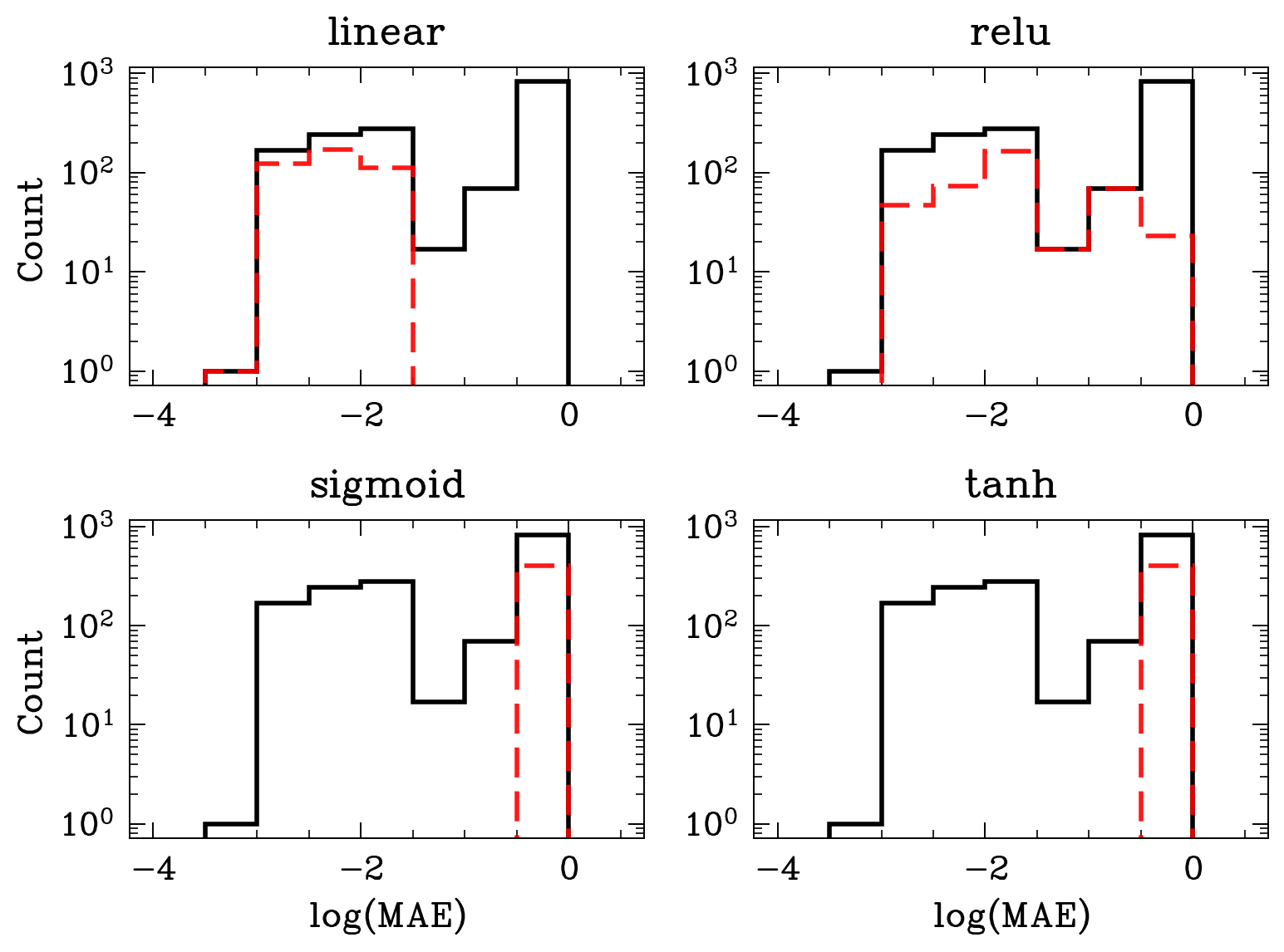}
	\caption{A diagnostic figure illustrating the results from \texttt{RSCV}. The figure consists of $N$ plots, where $N$ corresponds to the number of unique values for a given hyperparameter. Each plot features two histograms: the first histogram (solid black line) represents the entire sample, while the second (dashed red line) represents the subset with the specific value of the hyperparameter, shown above each histogram (in this case, the AF in the OL). The $x$-axis indicates the values of the chosen metric (here, the MAE averaged across all folds). The lower the value of the metric, the better the performance.
	}
	\label{fig:rscvolafmetrics}
\end{figure}

Figure~\ref{fig:rscvolafmetrics} depicts a single RSCV run, illustrating the performance of various AFs in the output layer (OL). The figure demonstrates that the \texttt{linear} AF accounts for the majority of favorable results (with lower values indicating better performance). We performed similar analyses across multiple ANN architectures; by architecture, we refer to both the structure of the ANN, encompassing the number of hidden layers (HLs) and the number of nodes in each HL, and the AFs employed in each layer.

Our hyperparameter search space included both shallow and deep ANNs, ranging from 2 to 10 HLs, with layers of varying size, from 16 to 2048 nodes\footnote{Although not strictly required, it is a common practice to use powers of two for the number of nodes.}. We evaluated the most commonly used AFs, such as \texttt{sigmoid} and \texttt{relu}, alongside less frequently used ones, including \texttt{linear}, \texttt{tanh}, and \texttt{elu}. While identifying a universally optimal architecture is practically infeasible and also pragmatically insignificant, our analysis of thousands of configurations revealed certain consistent patterns that considerably reduce the range of architectures likely to perform well for our specific problem:

\begin{itemize}
	\item The total number of nodes in the HLs should not be less than 128. Moreover, dividing a given number of nodes across multiple layers generally yields better performance. This trend is observed even for smaller OLs.
	\item Beyond a certain point, increasing the size of the ANN structure does not result in significant performance improvements. However, the size of the structure is directly correlated with computational time, both for training the model (assuming the same batch size and number of epochs) and for prediction.
	\item Given the importance of computational time in our work, we determined that ANNs composed of no fewer than 4 and no more than 6 HLs provide optimal performance. Furthermore, the number of nodes in each layer should neither be too small nor excessively large. By analyzing diagnostic figures, as presented in Figure~\ref{fig:rscvolafmetrics}, we found that using 256 to 1024 nodes per layer strikes the ideal balance. Models with too few nodes were unable to capture the full complexity of the problem, and the resulting light curves significantly deviated from the expected ones. On the other hand, models with an excessively high number of nodes began to suffer from overfitting, where the ANN performed well on the training set but produced inferior results on the validation set. Additionally, a several-fold increase in the time required for model predictions was also observed.
	\item The \texttt{sigmoid} and \texttt{tanh} AFs perform poorly when applied to the OL but show strong performance in the HLs. In contrast, the \texttt{linear} AF works well for the OLs but performs poorly in the HLs. \texttt{Relu} appears to be a versatile AF, functioning effectively in both the OLs and HLs.
	\item Among all AFs tested, \texttt{elu} showed the best performance when used in the first HL, slightly outperforming \texttt{tanh}. For the last HL, both \texttt{tanh} and \texttt{sigmoid} yielded the best overall performance, however \texttt{sigmoid} surpassed the other AFs in the remaining HLs.
	\item No significant differences were observed between the performance of the optimization algorithms \texttt{Adam}, \texttt{Adamax}, \texttt{Nadam}, and \texttt{RMSprop}. In contrast, \texttt{SGD}, \texttt{Adagrad}, and \texttt{Adadelta} underperformed. For subsequent tests, we chose to use the widely adopted \texttt{Adam} optimizer.
	\item Networks with non-decreasing node counts in the HLs consistently outperformed those where the number of nodes decreased at any point.
	\item The choice of batch size and number of epochs significantly affects an ANN's performance. Smaller batch sizes can enhance performance by allowing more frequent weight updates and reducing the risk of getting stuck in a local minimum. Conversely, larger batch sizes can expedite training, though they may result in suboptimal solutions and slower convergence, necessitating more epochs. Training time is directly proportional to the number of epochs; while it may be tempting to use a large number of epochs to improve the accuracy, this can lead to overfitting. The network may perform well on the training set but degrade significantly when applied to unseen data. In our tests, optimal batch sizes ranged from 500 to 1000, with the number of epochs varying from 1000 to several thousand. It is important to note that smaller batch sizes require more epochs.
\end{itemize}

To summarize this step, based on our analysis of various ANN architectures encompassing more than ten thousand cases, we determined that the most optimal architecture for a deep, fully connected ANN, designed for a regression problem where the goal is to predict a forward model (in this case, a list of fluxes) based on given parameters, consists of between 4 and 6 HLs arranged in a non-decreasing number of nodes, with each HL containing between 256 and 1024 nodes. The AF for the first HLs should be either \texttt{elu} or \texttt{tanh}, while for the remaining layers, we recommend using \texttt{sigmoid} or \texttt{tanh}. The AF in the OL should be \texttt{linear}. For the optimization algorithm, we recommend using \texttt{Adam}, although similar results may be achieved with \texttt{Adamax}, \texttt{Nadam}, or \texttt{RMSprop}. The recommended batch size is between 500 and 1000, and the number of epochs must be selected carefully to avoid overfitting. For large training sets comprising hundreds of thousands or millions of elements, the required number of epochs will likely range from several thousand to approximately 10,000.

\section{PHOEBE via AI} 

This work represents the second paper in a series applying AI methods to PHOEBE, building on the foundational work of \cite{prsa2008}, in which the authors introduced an AI model, EBAI, to estimate parameters based on time-series data. To distinguish the current ANN-based modeling process from the physical PHOEBE model, we have named it PHOEBAI, short for PHOEBE via AI.

Before applying an ANN to solve EB light curves, it is crucial to ensure that the ANN can accurately model ideal synthetic cases. If the ANN was to fail to produce reliable results for light curves generated under fully supervised conditions, where every orbital and physical parameter is known, it cannot be expected to perform reliably on actual observational data. Therefore, the ANN must meet a necessary condition, which in this context can be defined as its ability to model synthetic data within a required accuracy.

\subsection{Generating the Training Set}
\label{sec:training_set_generation}
To verify that the ANN can accurately model ideal synthetic light curves, we generated a comprehensive dataset using PHOEBE. The code accounts for various atmospheric effects, such as irradiation/reflection and limb and gravity darkening, and models stellar surfaces using the Roche approximation. Atmospheric parameters were computed using \cite{ck2004} model atmospheres for stars with effective temperatures ($T_{\mathrm{eff}}$) above 3500 K. For stars with $T_{\mathrm{eff}} \leq 6600$ K, we set the bolometric gravity darkening coefficient to 0.32 \citep{1967ZA.....65...89L} and the bolometric fraction of incident flux to 0.6 \citep{1990ApJ...356..613W}. For hotter stars, both coefficients were set to 1.0 \citep{2011A&A...529A..75C}. The main set of parameters varied in the generated dataset was sampled from specific initial distributions to capture a wide range of possible EB configurations: the orbital eccentricity \( e \) was drawn from a uniform distribution \(\mathcal{U}(0.0, 0.8)\); the cosine of the orbital inclination, \(\cos i\), was drawn from a uniform distribution \(\mathcal{U}(\cos 90^\circ, \cos 55^\circ)\); the argument of periastron, \(\omega\), was sampled from \(\mathcal{U}(0^\circ, 360^\circ)\); the sum of relative radii, \((R_1 + R_2)/a\), was drawn from \(\mathcal{U}(0.04, 0.75)\); and, the radius ratio, \(R_2/R_1\), as well as temperature ratio, \(T_2/T_1\), were sampled from log-normal distributions informed by mass–radius and mass–temperature relations.

We sampled the mass of the secondary, $M_2$, uniformly between 0.59 and 1.61$\,M_\odot$. For each sampled value of $M_2$, we determined central values of the secondary radius and effective temperature using mass-radius and mass-temperature relations computed based on Table 5 of \cite{2013ApJS..208....9P}. We then sampled secondary radius, $R_2$, and temperature, $T_2$, from log-normal distributions centered on logarithms of these central values, with standard deviations $0.075$ and $0.05$, respectively. This approach introduces realistic variations around the tabulated relations to simulate a diverse set of stellar properties, e.g., due to evolution on the main sequence. To maintain physical consistency, we enforced the conditions $R_2 \leq R_1$ and $T_2 \leq T_1$. The ratios of radii $R_2/R_1$ and effective temperatures $T_2/T_1$ were then computed from the sampled values. This method ensures that the properties of the secondary component are physically linked to its mass, rather than being sampled independently, resulting in more physically realistic binary system models.

Key physical characteristics of the primary were kept constant to provide a controlled environment for the ANN learning process. Specifically, the primary's mass was fixed at $M_1 = 1.61\,M_{\odot}$, radius at $R_1 = 1.728\,R_{\odot}$, and effective temperature at $T_1 = 7220$ K, corresponding to an F0-type star. This does not cause any loss of generality because additional networks can be trained for different primary stars.

For the sampled values of $e$ and $\omega$, the corresponding ranges for $e\sin\omega$ and $e\cos\omega$ were between $-0.8$ and $0.8$. To focus on systems with significant eclipsing events, we applied a filtering condition $\cos i < (R_1 + R_2)/[a(1-e)]$, based on the geometric requirement for eclipses to occur. Additionally, any systems where the amplitude of any eclipses was smaller than 0.01 were excluded, ensuring that only systems with significant, observable variations were included in the training set. A sample of light curves generated at this stage is presented in Figure~\ref{fig:pmglcs}.

\begin{figure}
	\centering
	\includegraphics[width=1.0\linewidth]{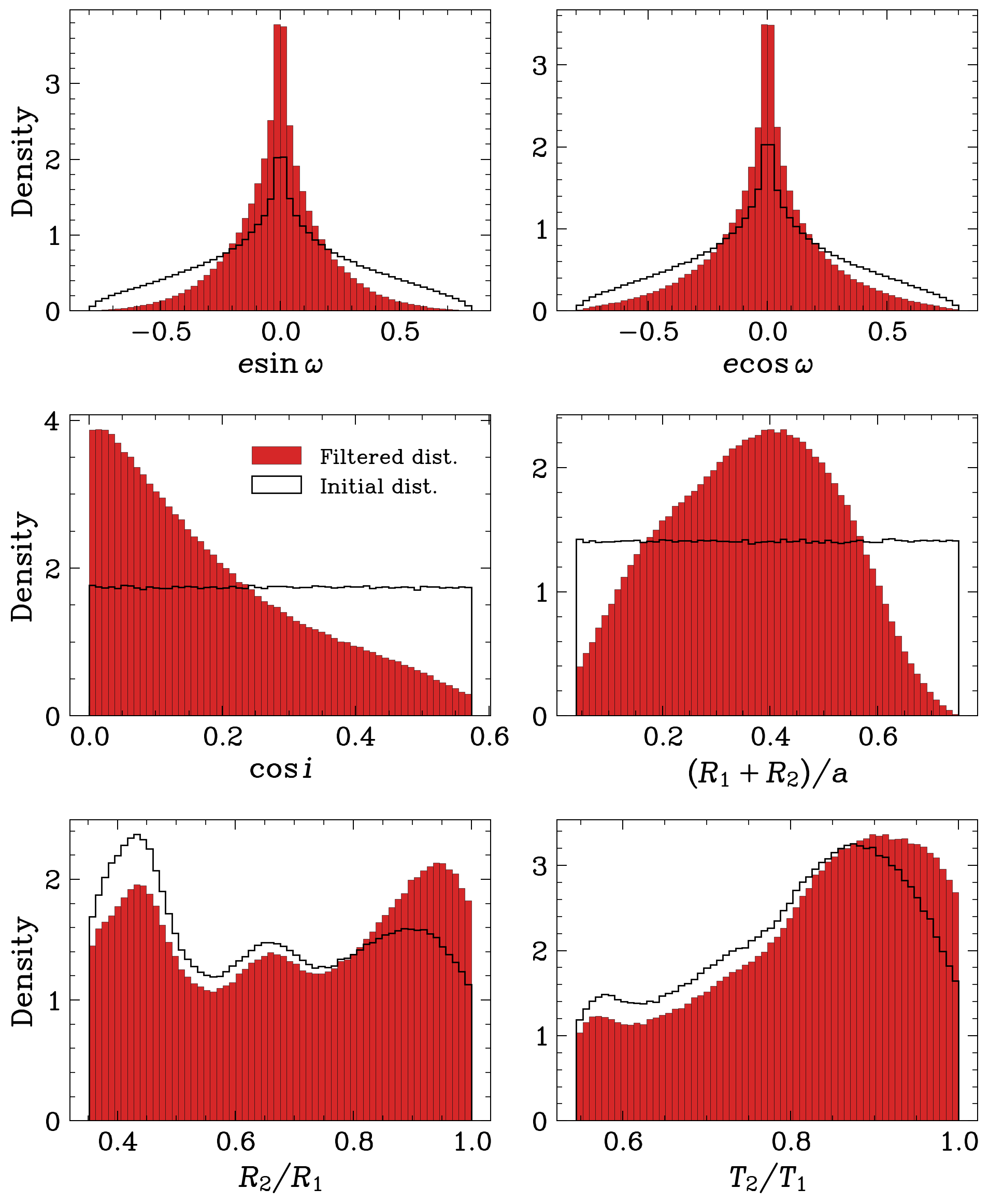}
	\caption{Density histograms comparing the initial and filtered distributions of the main set of parameters during the step of generating forward models using PHOEBE.
	}
	\label{fig:paramsposteriorhist}
\end{figure}

For certain combinations of randomly selected physical and orbital parameters, the system was not feasible. The most common issue, particularly in highly eccentric orbits or when the sum of the relative radii approached the semi-major axis, was mass transfer through the Roche lobe from the primary to the secondary (Roche lobe overflow, RLOF). Such cases were excluded from the final set of generated models. These factors, along with the mentioned limits on additional parameters, caused the filtered distribution of the main set of parameters to differ from the initial distribution. A comparison between these distributions is shown in Figure~\ref{fig:paramsposteriorhist}.

\subsection{Optimizers and samplers}
In the standard pipeline for solving an EB system with PHOEBE, there are three main steps: estimation, optimization, and sampling \citep{conroy2020}. During the estimation step, some parameters can be derived based on the geometric aspects of the light curve. Using this estimated solution, the next step involves optimizing additional parameters to refine the model fit to the data. This optimization step requires calculating multiple forward models (ranging from thousands to tens of thousands) and because of that can be highly time-consuming, taking up to a day per object. After optimization, MCMC sampling \citep{2010CAMCS...5...65G} is employed to explore the hyperspace around the optimized solution. This step helps detect correlations between parameters and estimate the heuristic uncertainties of the derived parameters. It is the most time-intensive part of the analysis, requiring the calculation of millions of forward models.

Solving an EB using an ANN follows similar steps, although the estimation and optimization steps may be merged into a single process. The computational time for a single forward model is about five orders of magnitude shorter using an ANN compared to PHOEBE. This allows for the optimization over a much larger parameter hyperspace. There is no need for parameter estimation, as we can efficiently cover a much broader hyperspace from the start.

In the optimization process, we use the differential evolution method from the \texttt{scipy.optimize} library \citep{scipy2020}. Generally, we optimize eight parameters: the main set of six parameters, along with a flux scale and phase shift. These additional parameters are not critical at this stage since all generated light curves are scaled to have the same baseline flux of 2.0, and the phase of the superior conjunction is always set to 0.0. Although the program does not receive this information as a prior, it consistently converges to a flux scale near 1.0 and a phase shift close to 0.0. To sufficiently cover the parameter hyperspace, a large population size (at least 50) is recommended. The objective function minimized by the optimizer is the \(\chi^2\) function, defined as:
\begin{displaymath}
	\chi^2(\mathbf{r}) = \sum_{k} \frac{\left[ y_k - y(x_k; \mathbf{r}) \right]^2}{\sigma_k^2},
\end{displaymath}
where \( y_k \) are the data values at the \( k \)-th point, \( y(x_k; \mathbf{r}) \) is the model prediction at the \( k \)-th point, evaluated using the parameters \( \mathbf{r} \), and \( \sigma_k \) are the uncertainties associated with each data point.

The process of incorporating PHOEBAI into predictions and comparing them to the data follows this approach: for a given parameter vector \(\mathbf{r}\), the program first normalizes it using a pre-fitted \texttt{scaler}\footnote{During ANN training, it is standard practice to normalize input data using a \texttt{scaler}. When using the ANN to predict a model, the input parameters must be normalized using the same scaling factors as those applied during the training and validation phases.} and then passes it to PHOEBAI to predict the light curve. The predicted light curve is interpolated using cubic interpolation over a phase range of \([-0.75, 0.25]\). The phase shift is applied to the input phase values, wrapping values outside this range back into it. The resulting interpolated model values are scaled by a flux scale factor.

The results obtained from the optimizer are passed to the sampler and used to set up the initial vector. The sampler has a structure similar to the optimizer, but instead of sampling the entire hyperspace and minimizing the error function, it performs local sampling around the optimized solution using MCMC. For this, we employ the \texttt{emcee} library in Python \citep{2013PASP..125..306F}. The model used in the sampling step differs only in that it no longer accounts for phase shift, although it still uses the flux scale factor. The scoring function used in the MCMC process is the log-probability, defined as:
\begin{displaymath}
	\log p(\mathbf{r} \mid \mathbf{D}) = \log p(\mathbf{D} \mid \mathbf{r}) + \log p(\mathbf{r}),
\end{displaymath}
where:
\begin{itemize}
	\item \( p(\mathbf{r} \mid \mathbf{D}) \) is the posterior probability of the parameters \( \mathbf{r} \) given the data \( \mathbf{D} \),
	\item \( p(\mathbf{D} \mid \mathbf{r}) \) is the likelihood function,
	\item \( p(\mathbf{r}) \) is the prior probability of the parameters.
\end{itemize}

Assuming Gaussian noise, the likelihood function is given by:
\begin{displaymath}
	\log p(\mathbf{D} \mid \mathbf{r}) = -\frac{1}{2} \sum_{k} \frac{\left[ y_k - y(x_k; \mathbf{r}) \right]^2}{\sigma_k^2},
\end{displaymath}
where \( y_k \), \( y(x_k; \mathbf{r}) \), and \( \sigma_k \) are as previously described for the \(\chi^2\) function.

The prior probability \( p(\mathbf{r}) \) reflects our prior knowledge of the parameters. In our case, we assumed uniform distributions within the min-max ranges of the generated models, ensuring that the ANN operates strictly in an interpolative manner rather than an extrapolative one.

\begin{figure*}[t]
	\centering
	\includegraphics[width=0.9\linewidth]{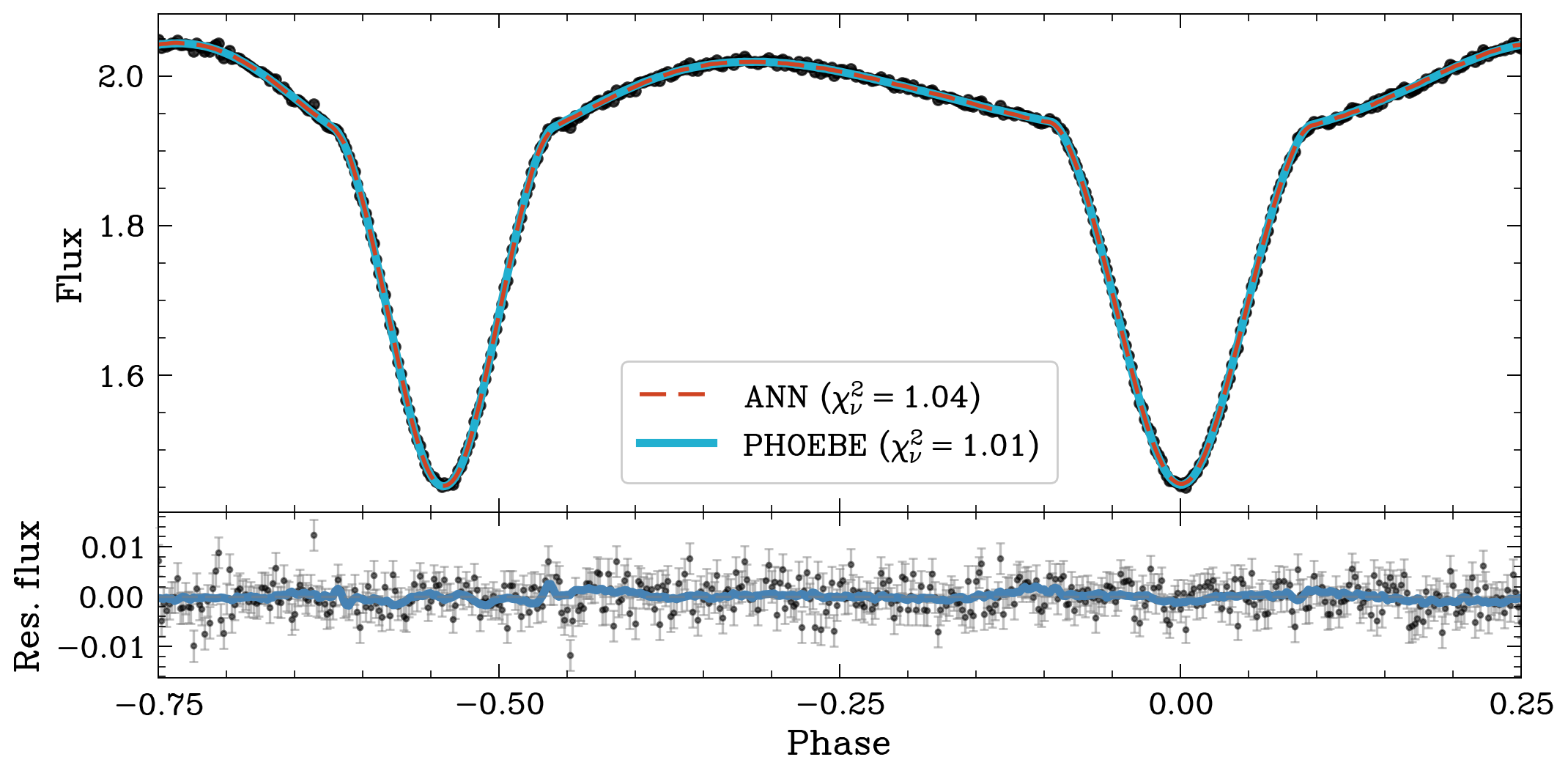}
	\caption{(\textbf{Top}) Phase-folded light curve of one synthetic dataset with fitted models. The dashed red line represents the model obtained using an ANN, while the solid blue line represents the model obtained with PHOEBE. (\textbf{Bottom}) Residuals after subtracting the ANN model fluxes from the observed data.}
	\label{fig:fvsp_model}
\end{figure*}

This step highlights a major advantage of using ANNs over PHOEBE for model generation. Depending on the size of the network and the dataset, sampling a single object takes only a few minutes using a single CPU core. In contrast, using PHOEBE requires significantly more time, often on the order of days to weeks, even when using high-performance computing clusters.

\begin{figure}
	\centering
	\includegraphics[width=1.0\linewidth]{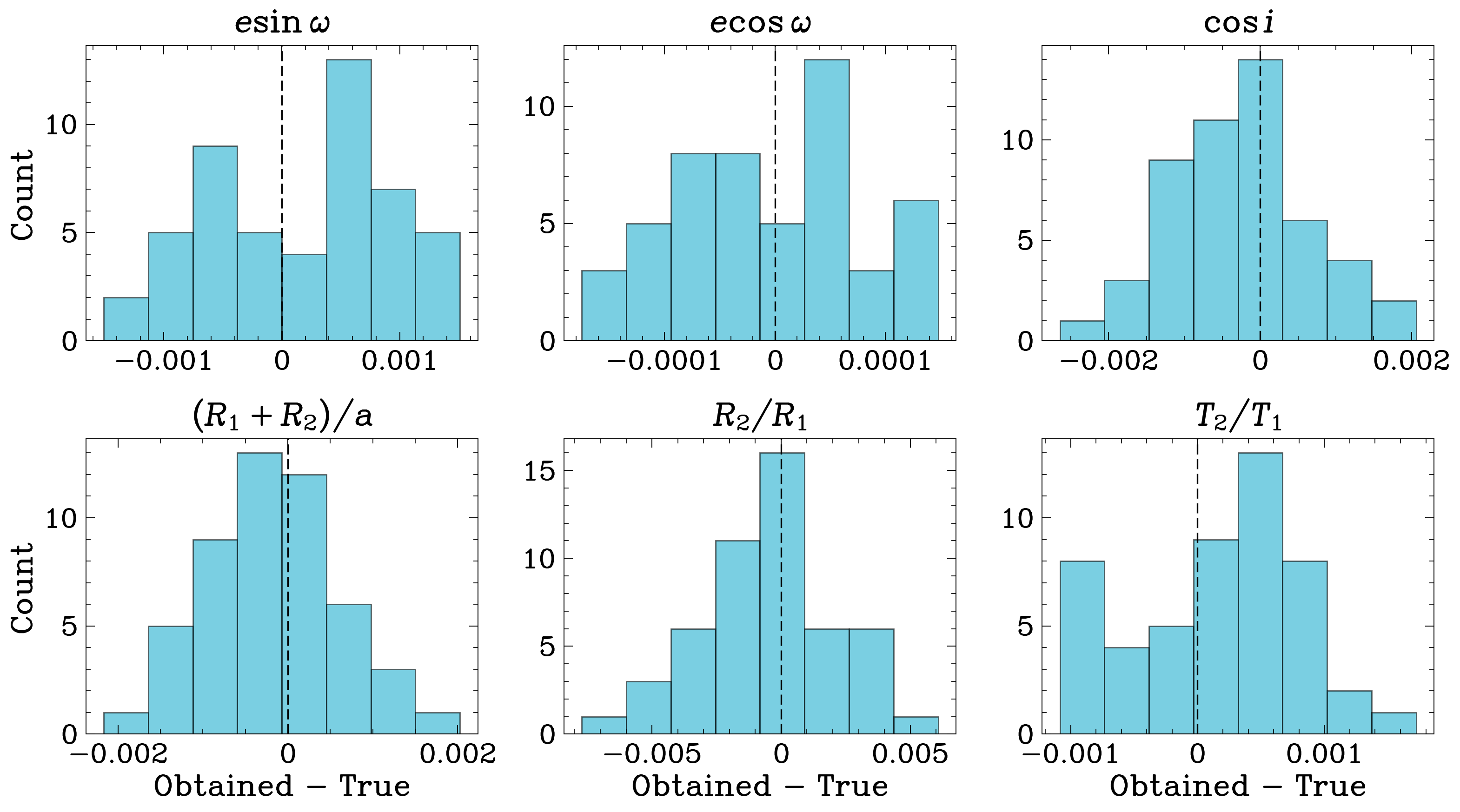}
	\caption{Histograms of the differences between the obtained parameter values and the true values derived using the bootstrapping method with PHOEBE. Black dashed lines denote the zero-point.
	}
	\label{fig:bootstrapingresphoebe}
\end{figure}

\begin{figure*}[t]
	\centering
	\includegraphics[width=0.9\linewidth]{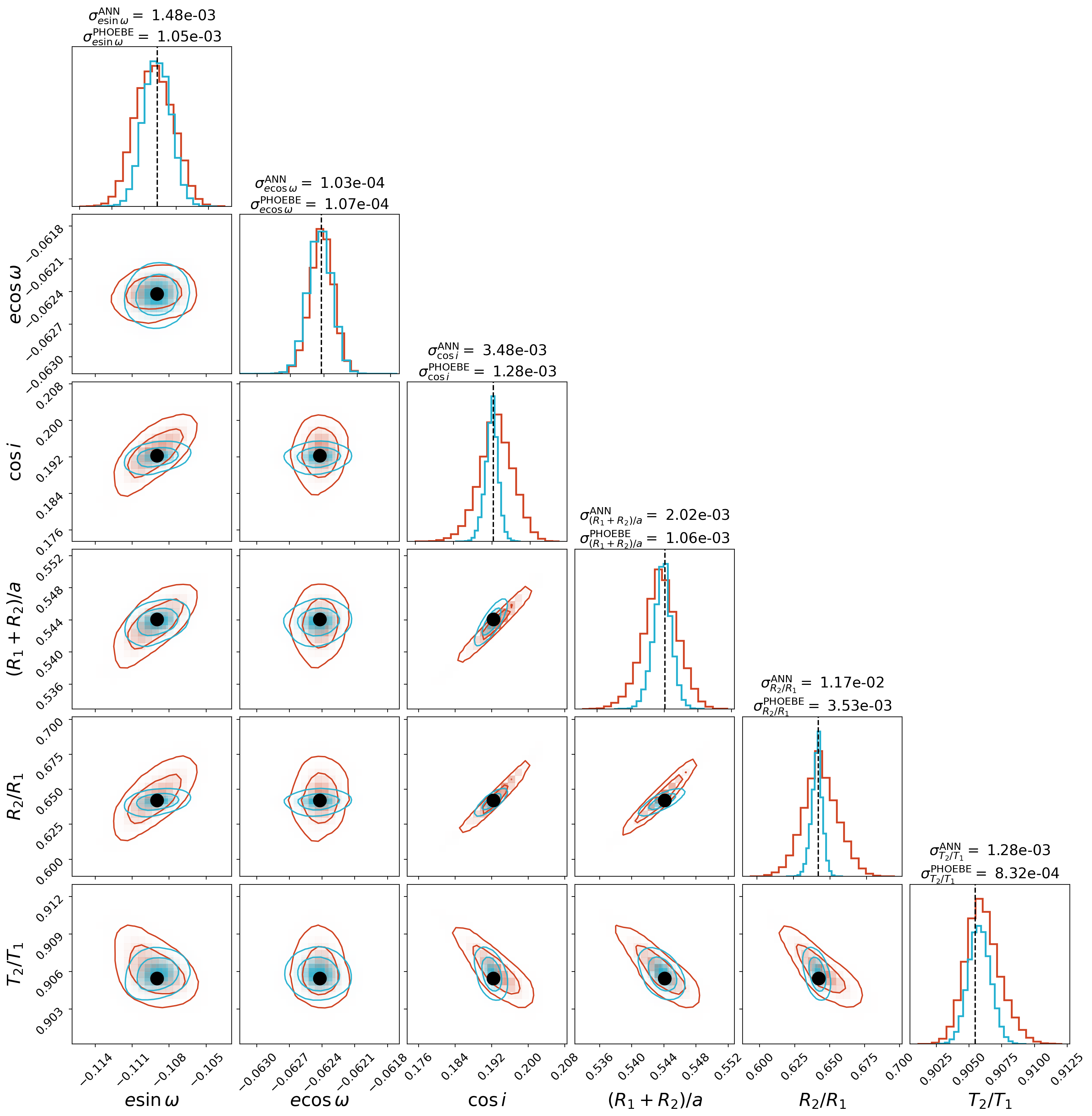}
	\caption{Corner plot of combined posterior distributions, comparing results from the bootstrap analysis for the final solution using ANNs (red) and PHOEBE (blue). Standard deviations are provided above each histogram. Black dots indicate the true parameter values used to generate the synthetic light curve.
	}
	\label{fig:fvsp_corner}
\end{figure*}

Using the same initial distributions as those used for generating the training set models, we generated a new synthetic light curve and conducted a full analysis using both PHOEBAI and PHOEBE. Figure~\ref{fig:fvsp_model} shows the phase-folded light curve with both fitted models and a plot of the residuals. Although both models are nearly identical, the PHOEBAI model exhibits small irregular oscillations in flux, while the PHOEBE model is smoother. Both models contain the same number of points (501).

To assess the robustness and accuracy of parameter determination for PHOEBE, we performed a bootstrap analysis. Specifically, a synthetic light curve was generated for the same system with 2000 data points, from which 501 points were randomly selected with replacement across 60 iterations. Each resampled light curve was modeled independently using PHOEBE. We analyzed the differences between the best-fit model parameters (corresponding to the highest log probability) and the true parameter values. Results are presented in Figure~\ref{fig:bootstrapingresphoebe}. The parameter determination accuracies were approximately $3 \times 10^{-3}$, $3 \times 10^{-4}$, $4.5 \times 10^{-3}$, $4 \times 10^{-3}$, $1.3 \times 10^{-2}$, and $2.5 \times 10^{-3}$ for the main set of parameters.

We also performed a modeling process for all bootstrap subsets using PHOEBAI. Figure~\ref{fig:fvsp_corner} shows the combined posterior distributions of the main six parameters obtained from both methods (blue for PHOEBE and red for PHOEBAI) across all bootstrap samples. The distributions of all parameters are generally similar between the two methods; however, some differences are noticeable. For example, the distributions for most parameters are slightly broader for PHOEBAI. The positions of the distributions from PHOEBE and PHOEBAI align well with the true parameter values (marked by black dots).

\subsection{Systematic and statistical errors}
\label{sec:sys_and_stat_errs}
\begin{figure*}[t]
	\centering
	\includegraphics[width=0.9\linewidth]{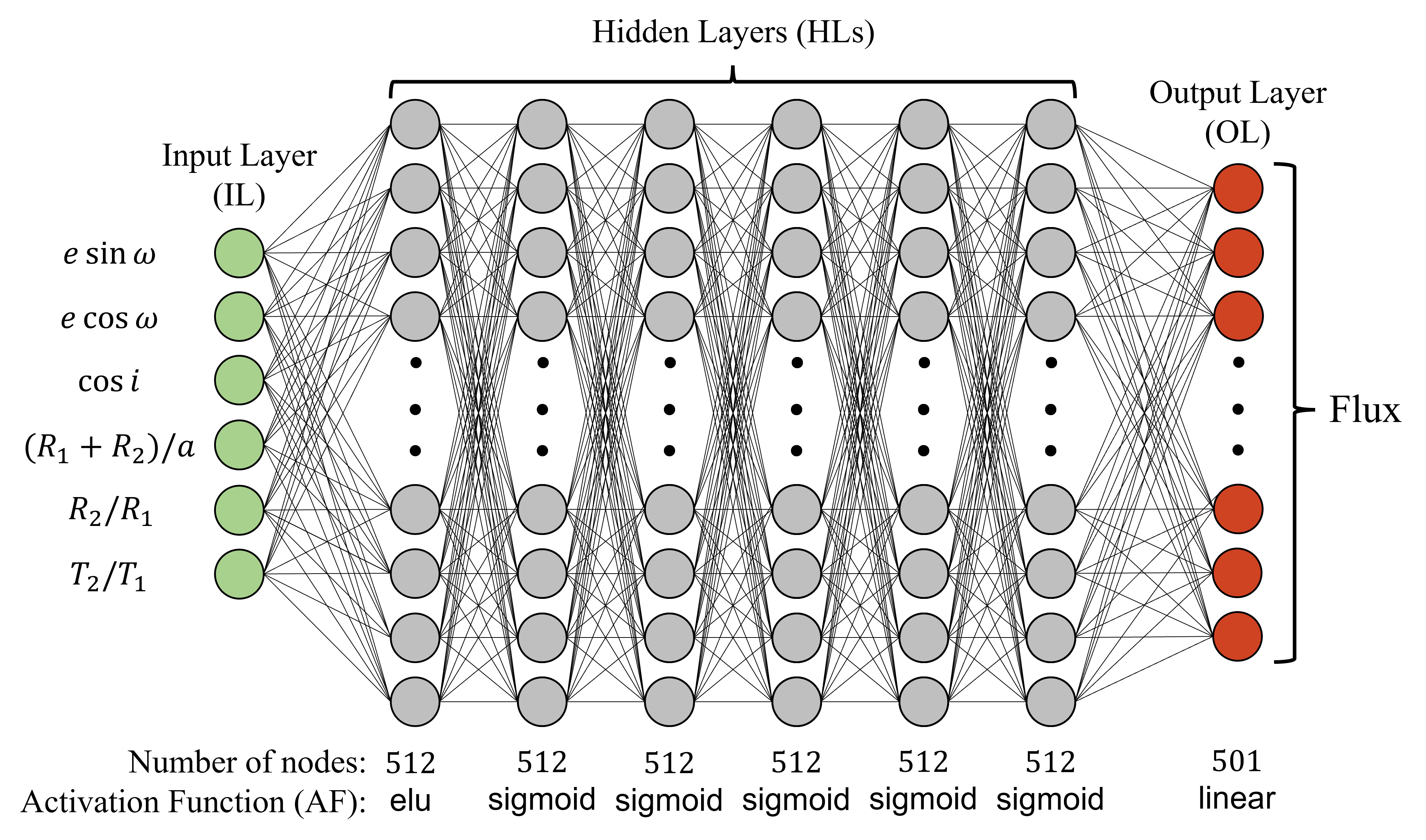}
	\caption{Architecture of the best-performing deep ANN based on cross-validation tests, using 1,000,000 objects for the training set and 250,000 for the validation set. This ANN contains six hidden layers (HLs), each with 512 nodes. The activation function (AF) for the first HL is \texttt{elu}, while \texttt{sigmoid} is used for the remaining HLs. The AF for the output layer (OL) is \texttt{linear}. The optimization algorithm used is \texttt{Adam}.
	}
	\label{fig:nnstructure}
\end{figure*}
To determine the limits of the ANN's applicability and estimate both systematic and statistical errors of the determined parameters, we prepared a separate dataset consisting of 1,250,000 binaries, with an additional 100,000 kept as an unseen sample (i.e., they were never used during the training of the ANN, either in the training set or validation set). Using this newly generated dataset, we validated the ANN architecture using the \texttt{GridSearchCV}\footnote{Unlike RSCV, GridSearchCV tests all combinations of parameters rather than just a random sample.} technique from the \texttt{sklearn} library. The final selected architecture achieved close to the lowest values for various loss functions and was also one of the fastest in the tested set. The details of the architecture of this ANN are shown in Figure~\ref{fig:nnstructure}.

Next, we conducted tests to examine how various aspects of data quality impact the ANN's performance and the precision of the solutions obtained using PHOEBAI. First, we tested whether the ANN is data dependent by performing cross-validation. Then, using synthetic data with various levels of injected noise, we analyzed how this noise affects the accuracy of the obtained parameters and posterior distributions. In a subsequent test, we examined the relationship between the accuracy of the determined parameters and their values. Finally, we evaluated how the number of available data points influences the spread of the posterior distributions and the accuracy of the solution.

\subsubsection{Cross validation check}
\label{sec:CV_check}
The primary objective of this cross-validation test was to verify whether the performance of our ANN is independent of the specific datasets used for training and validation. We conducted cross-validation by splitting the entire dataset into five equal parts (validation sets), with the remaining portion used as the training set. This process involved training five different ANNs, each using a different validation set. Two critical checks were performed: first, comparing learning curves across different models to ensure consistent behavior during training and validation; and second, comparing the posterior distributions obtained with the PHOEBAI sampler for different light curves to assess the stability and accuracy of the parameter estimation.

\begin{figure}
	\centering
	\includegraphics[width=1.0\linewidth]{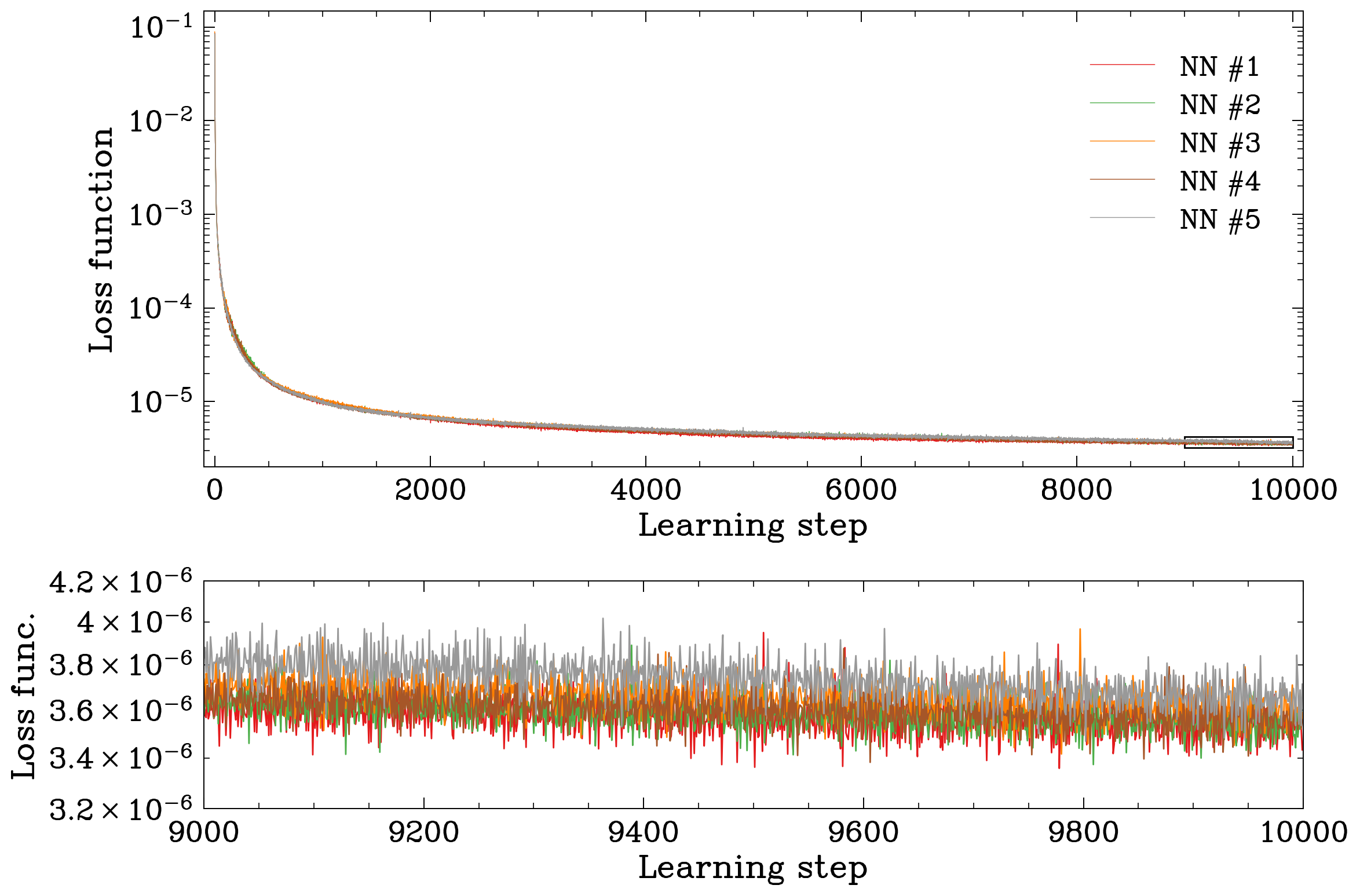}
	\caption{(\textbf{Top}) Learning curves for an ANN trained on five different subsets of the training set as part of a cross-validation test. In each variant, the training set consisted of one million entries, and the validation set of 250,000. In each variant, the validation set consisted of different objects, and the training set of the remaining part. (\textbf{Bottom}) Differences between the various learning curves become noticeable only when zooming in on the last 1000 steps of the training session.
	}
	\label{fig:cvlearningcurvetraining}
\end{figure}

The learning curves for both training and validation datasets are consistent across all models, with no significant deviations, indicating that the ANN's performance is stable and independent of the specific training and validation sets used (Fig.~\ref{fig:cvlearningcurvetraining}). For each training session, we used the same batch size and number of epochs.

\begin{figure}
	\centering
	\includegraphics[width=1.0\linewidth]{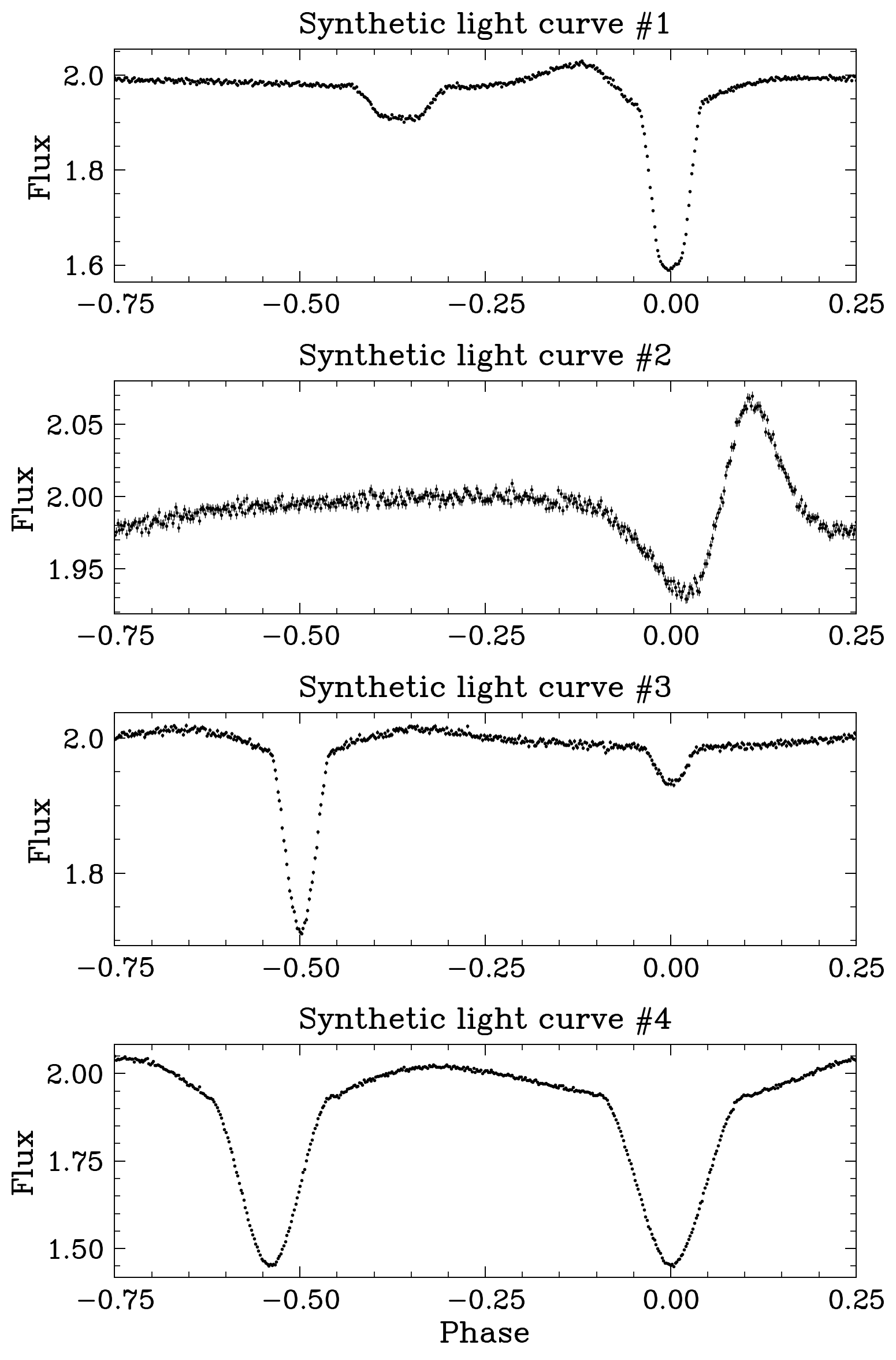}
	\caption{Phase-folded synthetic light curves that were used to evaluate the stability and performance of ANNs during cross-validation.}
	\label{fig:plotlcscv}
\end{figure}

\begin{figure*}
	\centering
	\includegraphics[width=0.7\linewidth]{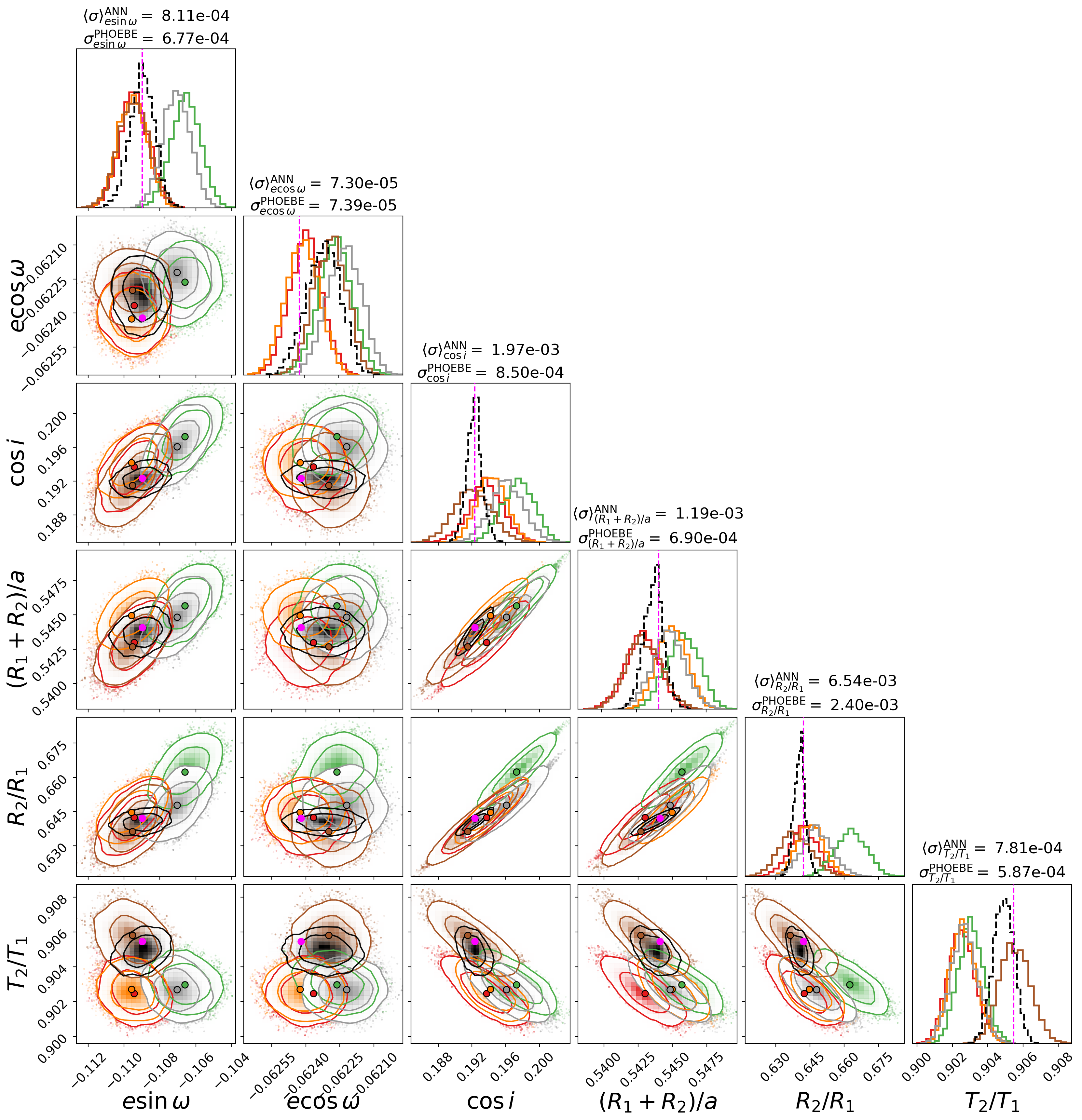}
	\includegraphics[width=0.8\linewidth]{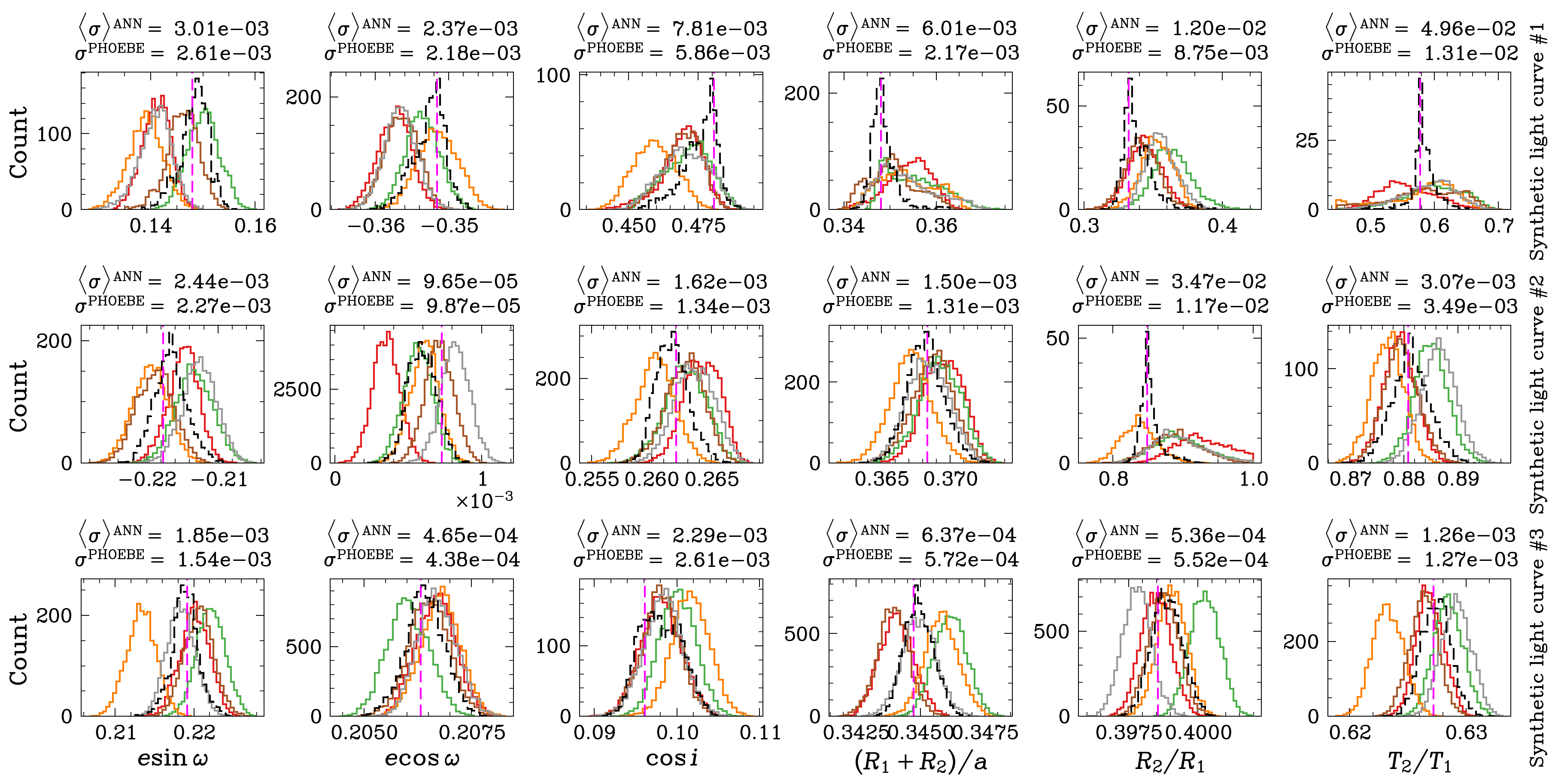}
	\caption{(\textbf{Top}) Corner plot with posterior distributions after sampling the solutions found by all five ANNs during the cross-validation test (colors match those used for learning curves in Fig.~\ref{fig:cvlearningcurvetraining}) and using PHOEBE (black). Here we present results for synthetic light curve \#4. True values are marked with magenta dashed lines on the histograms and magenta dots on the 2D plots. Above the histograms, we provide the average standard deviation of the distributions for a given parameter \(\mathcal{X}\) over all five ANNs \(\bigl\langle \sigma \bigr\rangle_{\mathcal{X}}^{\mathrm{ANN}}\) and the standard deviation of the posterior distribution obtained with PHOEBE \(\sigma_{\mathcal{X}}^{\mathrm{PHOEBE}}\). (\textbf{Bottom}) Histograms of the posterior distributions of the fitting parameters for the remaining three synthetic light curves. Line colors are the same as those used in the corner plot.
	}
	\label{fig:corner_CV}
\end{figure*}

The second check involved analyzing the posterior distributions obtained after modeling four synthetic light curves, presented in Figure~\ref{fig:plotlcscv}. In parallel, we also conducted a sampling procedure with PHOEBE for these synthetic light curves. To present these results concisely, we include a corner plot comparing PHOEBAI with PHOEBE results (upper panel in Fig.~\ref{fig:corner_CV}) and another figure with histograms of the posterior distributions for the remaining light curves (lower panel in Fig.~\ref{fig:corner_CV}). Most of the posterior distributions for the parameters lie within the $3\sigma$ range from the peak, indicating reliable parameter estimation across different datasets. The distributions from PHOEBE generally aligned well with the PHOEBAI results, although in some cases, PHOEBE produced narrower distributions. Notably, the $e\cos\omega$ parameter consistently showed a narrower distribution compared to $e\sin\omega$. However, one significant discrepancy was observed, where the PHOEBE distribution for $e\cos\omega$ deviated significantly from the true value (see middle row in lower panel of Fig.~\ref{fig:corner_CV}).

\subsubsection{Impact of data uncertainties}
\label{sec:data_uncertain}
For this test, we prepared six new objects, each with 15 light curves, varying the injected noise levels in the flux. As a reminder, all the generated light curves have a base flux of 2.0, and each light curve contained 501 evenly spaced phase points. We applied white noise with $\sigma$ values taken from the set ranging from $10^{-6}$ to $5 \times 10^{-2}$, equally spaced on a logarithmic scale. For each light curve, we performed the modeling process using PHOEBAI. The final parameter values were taken from the model with the highest log probability.

\begin{figure*}
	\centering
	\includegraphics[width=0.85\linewidth]{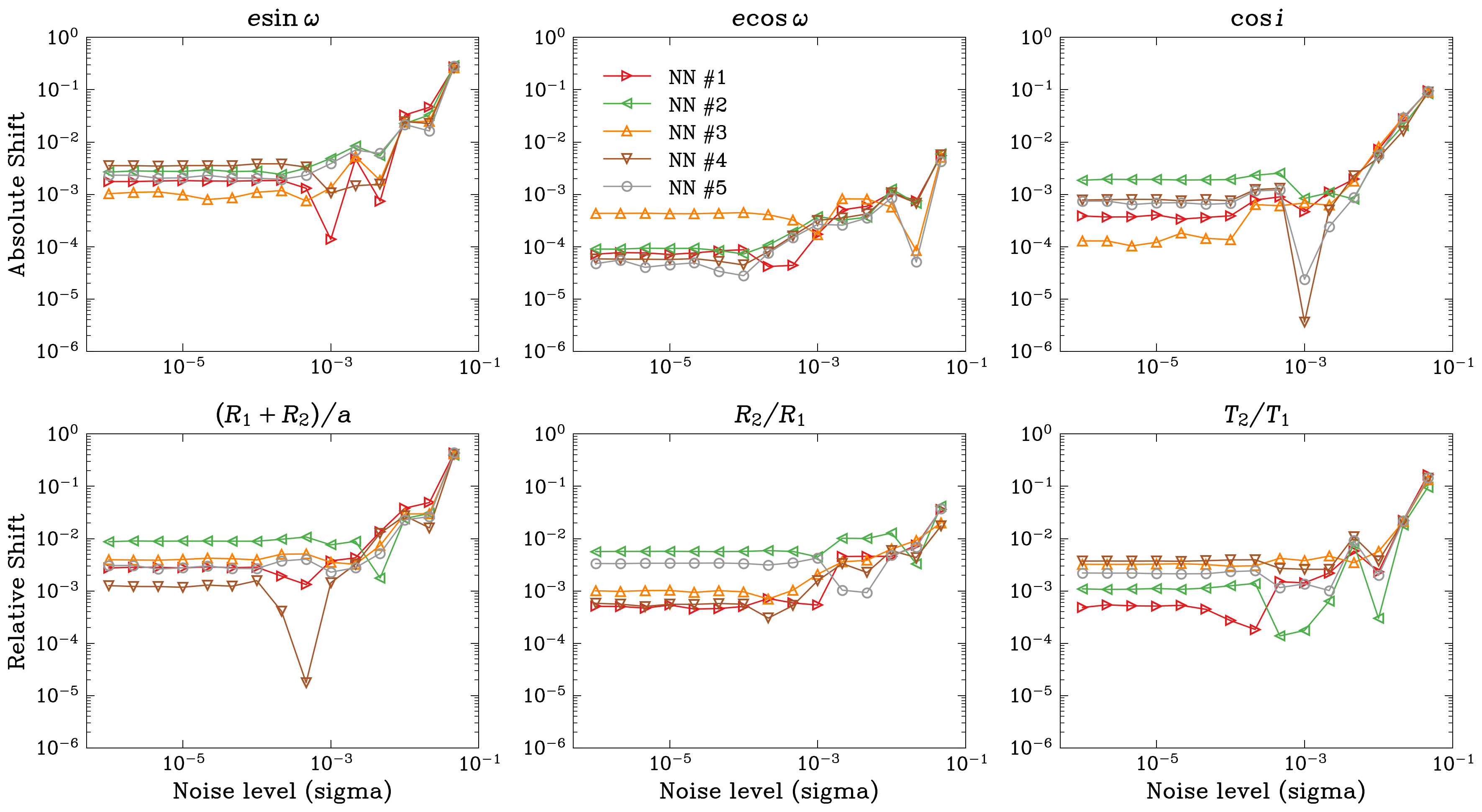}
	\textcolor{white}{\rule[0.5ex]{\linewidth}{0.5pt}}
	\includegraphics[width=0.85\linewidth]{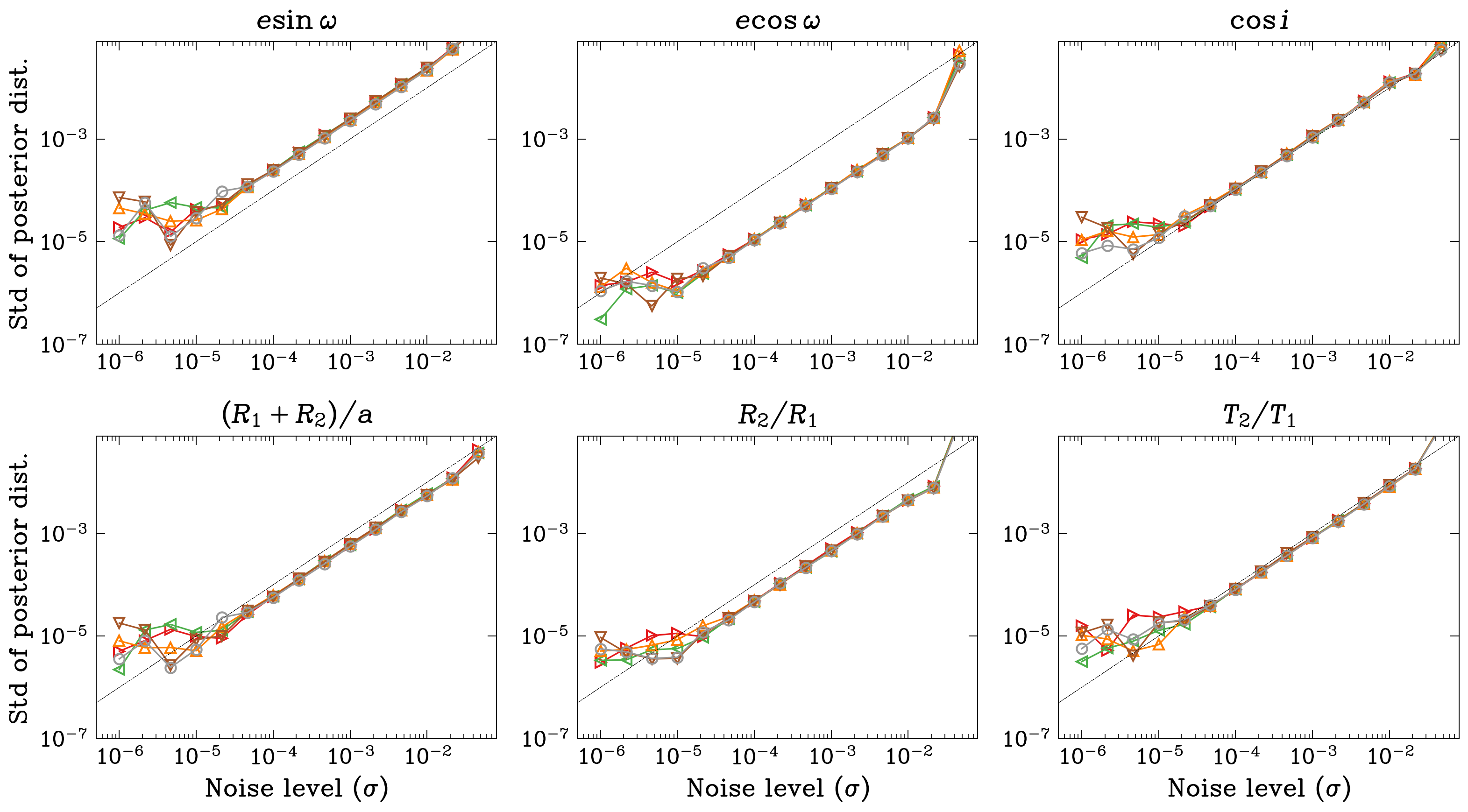}
	\caption{The relationship between the injected noise level (Gaussian, with the x-axis representing the noise \(\sigma\)) and the parameter shifts and scatter for one synthetic system. (\textbf{Top}) Parameter shift, defined as the absolute or relative difference between the true parameter value and the value determined from the model with the highest log probability, plotted against \(\sigma\). The five curves correspond to results from five different ANNs. (\textbf{Bottom}) Standard deviation of the posterior distribution for each parameter. A black reference line represents \(y = x\).
	}
	\label{fig:sigma_vs_shift_scatter}
\end{figure*}

We expected lower noise levels to result in more accurate parameter values, which turned out to be true, but only for significant noise levels, as shown in the top panel of Figure~\ref{fig:sigma_vs_shift_scatter}. The intrinsic noise in the ANN prevents parameter values from being determined with accuracy greater than 0.5\% for most parameters (0.01\% for \(e\cos\omega\)).

The bottom panel of Figure~\ref{fig:sigma_vs_shift_scatter} shows the relationship between noise level and the standard deviation of the posterior distribution for each parameter. As expected, higher noise levels resulted in a greater spread. However, the consistency of distributions across different networks and the strong linear correlation are surprising. It is evident that the lines are nearly parallel to the \(y = x\) line.

\subsubsection{The impact of parameter values}
\label{sec:params_value}
\begin{figure*}
	\centering
	\includegraphics[width=1.0\linewidth]{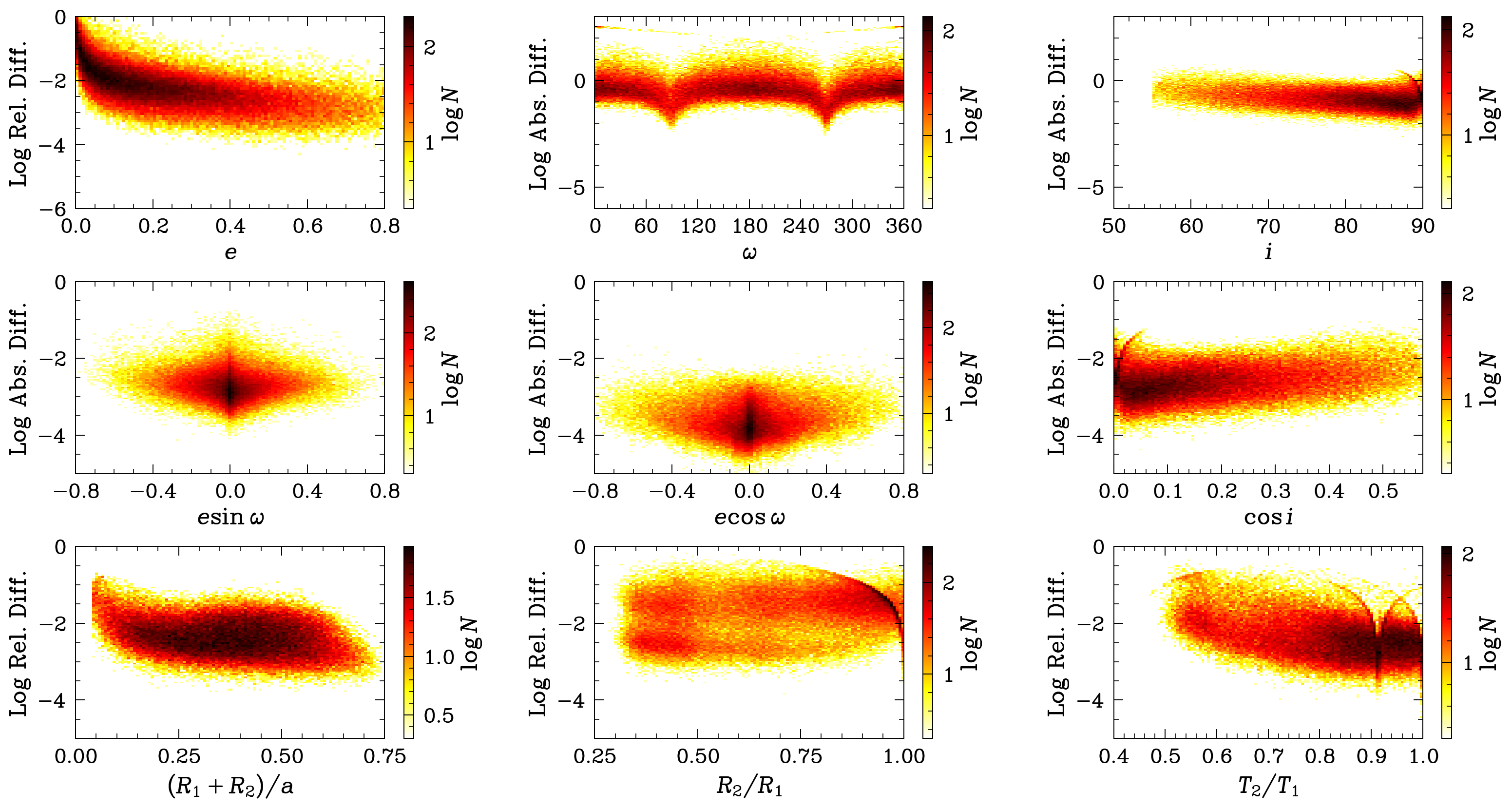}
	\caption{2D histograms showing the differences between the true parameter values and the highest log-probability estimates. The 2D histograms represent the average parameter shifts and scatter obtained using all five ANNs trained during the cross-validation test. The top row displays the orbital parameters \(e\), \(\omega\), and \(i\), plotted against their respective absolute or relative differences. The second and third rows present the main set of parameters: \(e\sin\omega\), \(e\cos\omega\), \(\cos i\), \((R_1 + R_2)/a\), \(R_2/R_1\), and \(T_2/T_1\). The color scale represents the logarithm of the number of occurrences in each bin (\(\log N\)).
	}
	\label{fig:shifts2dcombined}
\end{figure*}

During model generation for a new dataset, we kept 100{,}000 objects separate from the training sessions. We performed a sampling procedure on all light curves from this ``unseen'' dataset to evaluate generalizability. Each light curve contained 501 data points, as before. However, to make the data more realistic, we added noise to the flux, representing the typical relative noise level for TESS EBs at 0.00155\footnote{We selected \(\sim3000\) EBs from the Villanova TESS EBs catalog \citep{prsa2022}, calculated the average uncertainty for each object, and used the median of those averages.}. As in previous steps, the final parameter values were taken from the model with the highest log probability. The results for this step are presented as 2D histograms in Figure~\ref{fig:shifts2dcombined}.

The plots reveal certain patterns, glitches, and nonobvious trends that require further discussion:
\begin{itemize}
	\item $e$ -- The most pronounced feature of the plot is the increasing relative shift with decreasing eccentricity. This behavior is expected, as $e$ has a minimum value of 0, causing relative values to surge near this point. Another contributing factor is that, for nearly circular orbits, the argument of periastron is poorly determined. Since the model solves for $e\sin\omega$ and $e\cos\omega$, this limits the precise determination of $e$ as well.
	
	\item $\omega$ -- For most systems, the absolute error is below 1\degree, but there is a small group with much larger shifts. This is due to the fact that, for small eccentricities, $\omega$ becomes undefined, making any outcome possible. The other notable features on the plot are the distinct minima near $\omega = 90\degree$ and $\omega = 270\degree$. As seen on the $e\cos\omega$ plot, this parameter can be determined with higher accuracy than $e\sin\omega$, because it relates to the distance between the two eclipses. At $\omega = 90\degree$ and $\omega = 270\degree$, the derivative of $e\cos\omega$ reaches its extremes, making the parameter most sensitive to changes in $\omega$. This means $\omega$ can be determined with higher accuracy at these points, resulting in the lowest absolute shifts.
	
	\item $i$ and $\cos i$ -- Both plots feature a minimum occurring just before $i = 90\degree$. This is because, for a system with given values of $R_2/R_1$ and $(R_1 + R_2)/a$, a specific range of inclinations will result in total eclipses. The critical inclination separating total from partial eclipses is significant, as it induces a distinctive flattening at minimum brightness in the light curve. At higher inclinations, the impact on light-curve shape diminishes. The plots also show arcs near 90\degree. These represent systems for which the network yielded an edge value, in this case, 90\degree. Such cases occur when the program fails to find a proper solution (for example, if the light curve amplitude is too small relative to the noise) or when a given parameter is strongly correlated with another parameter.
	
	\item $e\sin\omega$ and $e\cos\omega$ -- Both distributions exhibit a triangular shape, which is more pronounced for $e\cos\omega$. This pattern aligns with the plot for $e$, where the relative shift stabilizes around $e = 0.3$, implying an increase in absolute error. Precision decreases as $e\sin\omega$ and $e\cos\omega$ approach zero due to fewer systems in this parameter space. Nevertheless, even at high eccentricities, accuracy remains acceptable, reaching $10^{-3}$ in the worst cases.
	
	\item $\rsum$ -- The plot shows a mostly flat distribution. However, similar to $e$, the relative deviation increases sharply as $(R_1 + R_2)/a$ approaches small values. This decrease in accuracy is also influenced by the fact that, as the value of this parameter becomes smaller, the number of points in the eclipses decreases because they last for a shorter time relative to the orbital period.

\begin{figure*}
	\centering
	\includegraphics[width=0.8\linewidth]{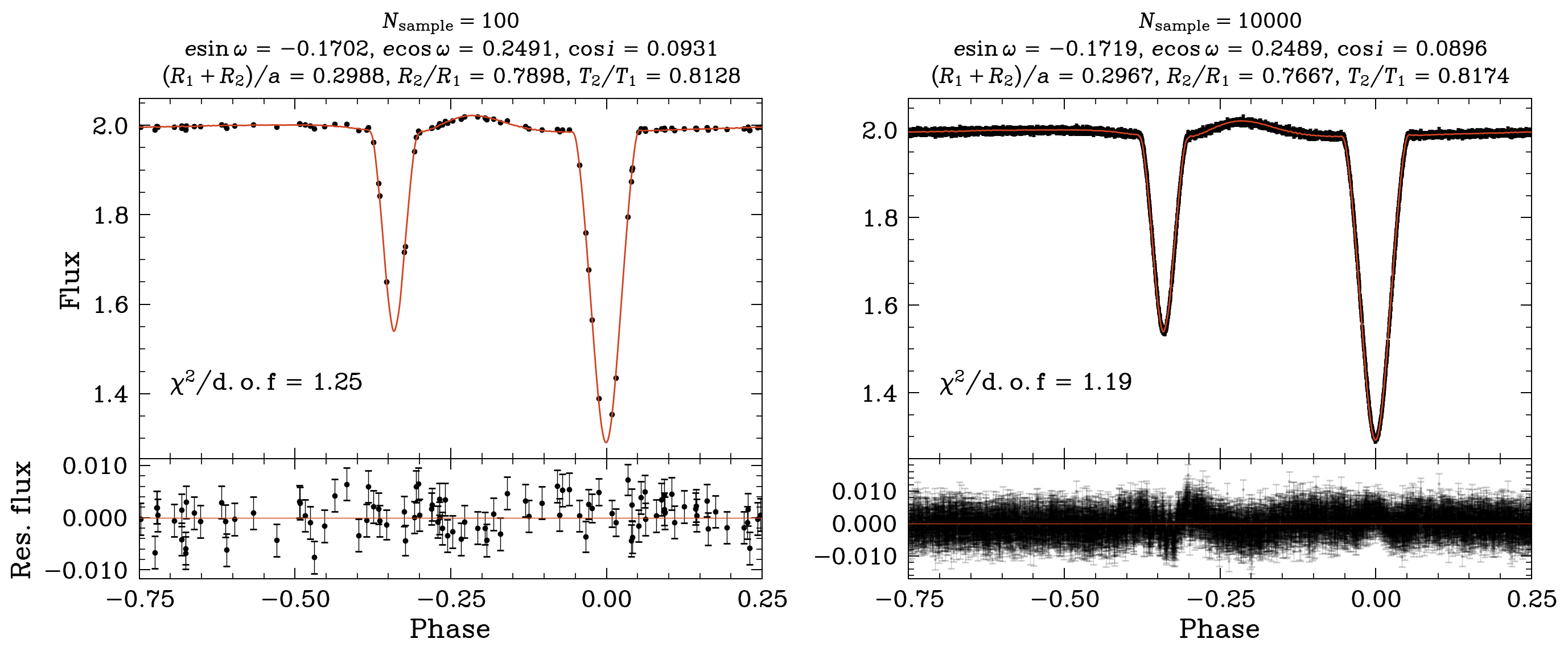}
	\caption{Phase-folded light curves of one synthetic object plotted alongside their respective PHOEBAI-fitted models. The left-hand plot shows the light curve with 100 data points, while the right-hand plot shows the same curve with 10{,}000 data points. Despite the large difference in sample size, the obtained parameters are consistent with the true values, which are: \(e\sin\omega = -0.1726\); \(e\cos\omega = 0.2489\); \(\cos i = 0.0871\); \((R_1+R_2)/a = 0.2967\); \(R_2/R_1 = 0.7485\); and \(T_2/T_1 = 0.8195\).
	}
	\label{fig:sslcmodels}
\end{figure*}
	
    \item $\rratio$ -- This plot displays a characteristic split into two groups, with a horizontal separation around $-2$. The lower group primarily consists of systems exhibiting total eclipses. In such cases, the radius ratio significantly impacts light curve shape, enabling a more accurate estimation of $\rratio$. The second distinct feature is an arc similar to those seen in the plots of $i$ and $\cos i$. It results from the overrepresentation of systems for which the network yielded an edge value of the parameter.

\item $T_2/T_1$ -- The plot shows some arcs, which appear as straight lines when the $y$-axis is not in logarithmic scale. The lines starting at the edges, corresponding to $T_2/T_1 = 0.6$ and $T_2/T_1 = 1.0$ (which represent $T_2 = 4200\,\mathrm{K}$ and $T_2 = 7000\,\mathrm{K}$), are due to the optimizer and sampler reaching their imposed limits for this parameter. A similar situation occurs for the second edge. The starting point at $T_2/T_1 \approx 0.94$ ($T_2 = 6600\,\mathrm{K}$) for the second pair of lines is related to the limit where atmospheric parameters change (the transition from convective to radiative envelopes in main-sequence stars) while generating the models. The program identifies this value as a convenient local minimum, causing a small group of EBs to converge on a temperature ratio close to this value instead of slightly above or below it.
\end{itemize}

This test revealed two encouraging findings. First, there were no spurious or unaccounted-for systematic effects, and all correlations between the achieved accuracy and specific parameter values, as well as any patterns observable in the plots, can be explained. Second, conducting this test required modeling light curves for as many as 100{,}000 EBs, demonstrating that ANN can be successfully applied even to such a large database.
\newline
\subsubsection{Impact of data size}
\label{sec:data_size}
In this test, we examined how the number of data points impacts the sampler performance. We generated another set of light curves and created multiple versions with different sample sizes from the set \{100, 250, 500, 1000, 2500, 5000, 10{,}000, 25{,}000, 50{,}000, 100{,}000\}. We generated 10 objects with randomly selected parameters and saved 500{,}000 data points for each object. Then, we drew random phases to achieve the desired sample size. As in the previous steps, we included a noise level of 0.00155 in the fluxes.

\begin{figure}
	\centering
	\includegraphics[width=1.0\linewidth]{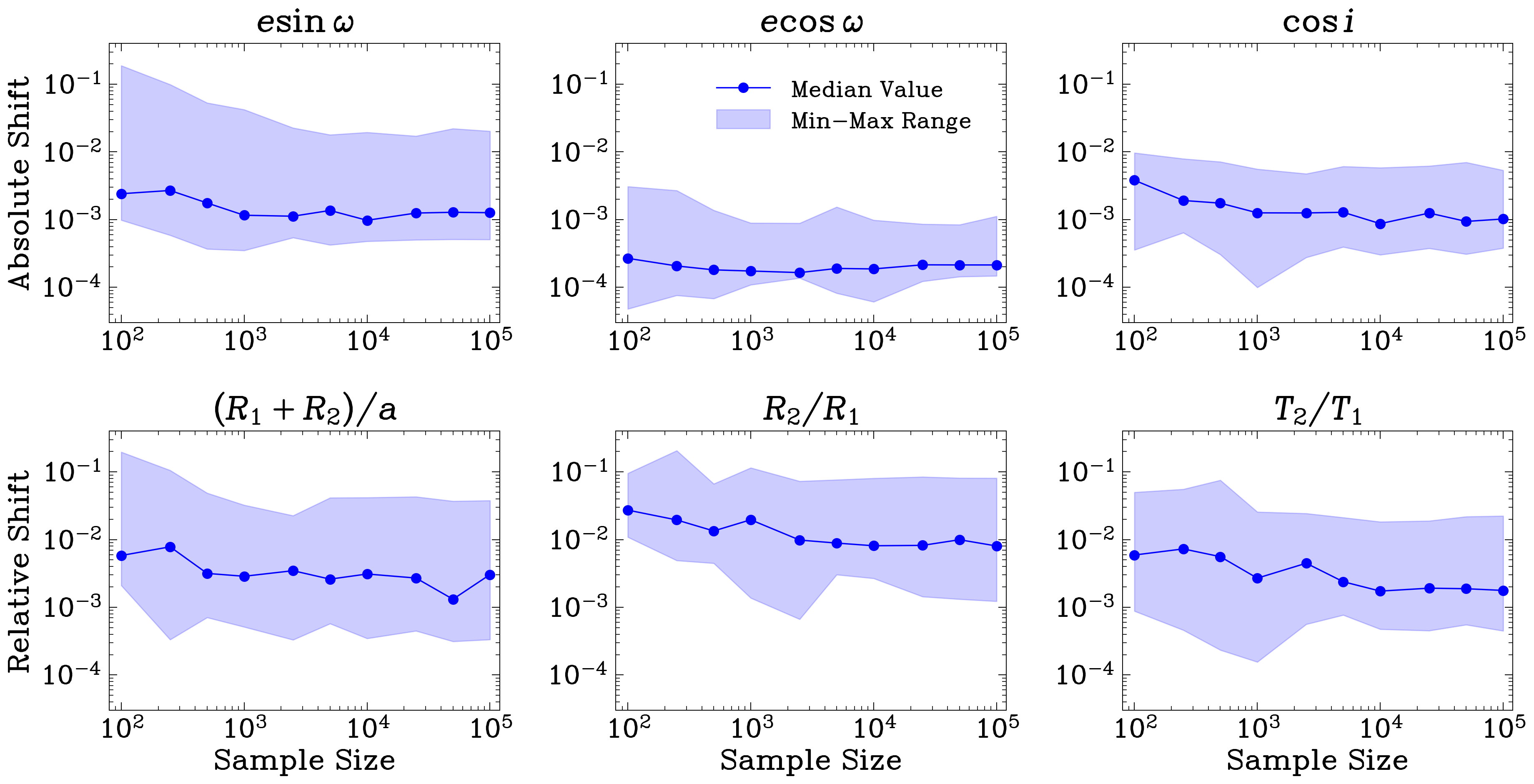}
	\textcolor{white}{\rule[0.5ex]{\linewidth}{0.5pt}}
	\includegraphics[width=1.0\linewidth]{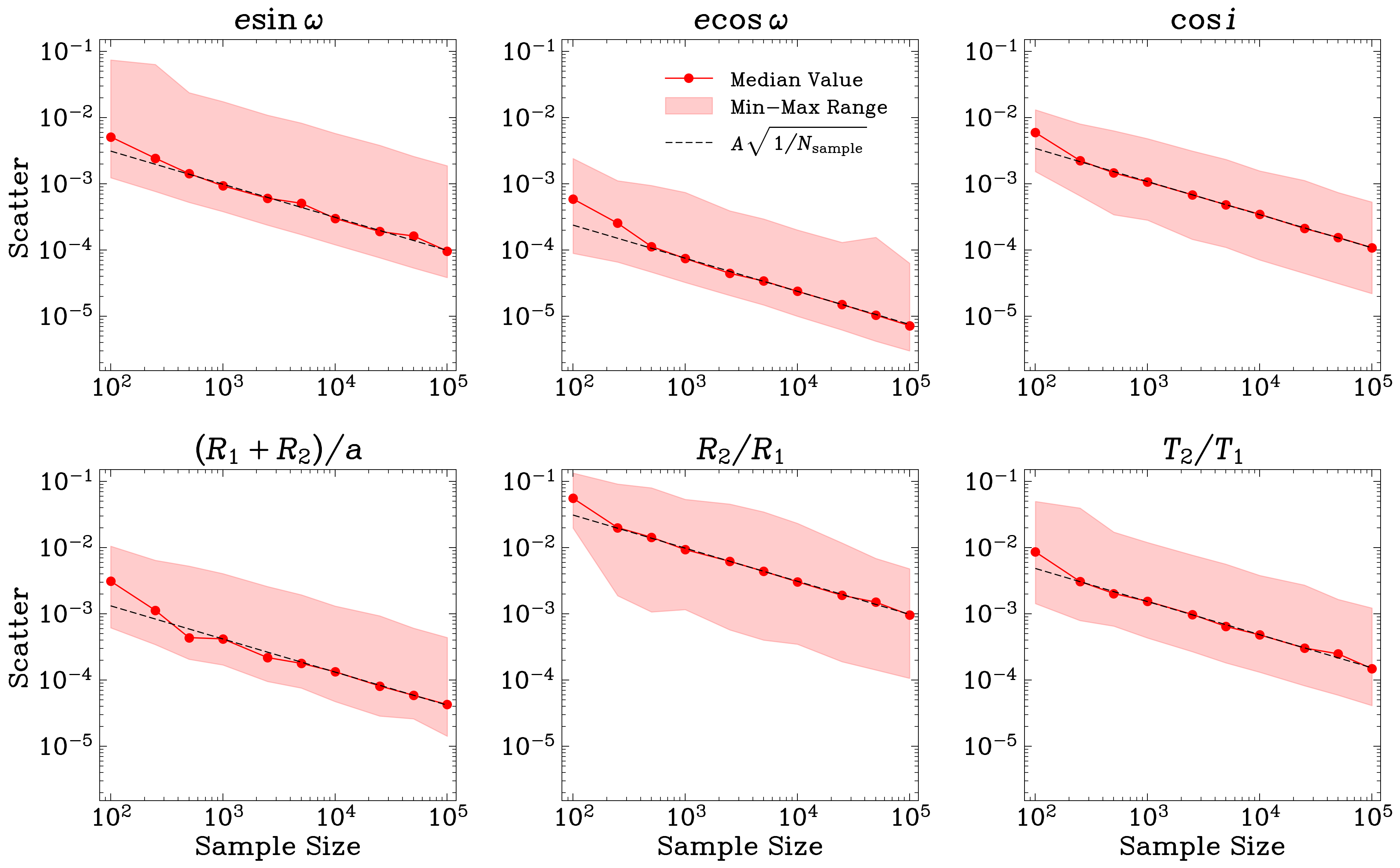}
	\caption{(\textbf{Top}) Median and min–max range of absolute and relative shifts of the parameters obtained using PHOEBAI for a group of 10 synthetic light curves, plotted against the sample size. For each sample size, all shifts are averaged over the five ANNs trained during the cross-validation test. The same range is used for all axes. (\textbf{Bottom}) Median and min–max range of the posterior distribution scatter (standard deviation) for the main parameters after MCMC simulations using PHOEBAI. The median line represents the average value for all 10 synthetic systems, averaged over all five ANNs. Black dashed lines represent the theoretical fit showing that scatter is inversely proportional to the square root of the number of points.}
	\label{fig:sample_size_SnS}
\end{figure}

Figure~\ref{fig:sslcmodels} shows a comparison of light curves and fitted models for a single object sampled with 100 and 10{,}000 data points. Despite the drastically different amount of data, the obtained parameters remain consistent. Figure~\ref{fig:sample_size_SnS} presents the combined results. For each sample size and each object, the median shift or scatter was calculated across all five ANNs. The plots show the minimal, maximal, and median values over all 10 objects, allowing us to visualize trends and result variability. The axes display absolute shifts for $e\cos\omega$, $e\sin\omega$, $\cos i$ and relative shifts for $(R_1 + R_2)/a$, $R_2/R_1$, and $T_2/T_1$. The shifts represent the difference between the true parameter values and the model values with the highest log probability. The scatter in the lower panel of Figure~\ref{fig:sample_size_SnS} represents the standard deviations of the posterior distributions.

Interestingly, the number of data points does not significantly affect the accuracy of the parameter determination. In the top panel of Figure~\ref{fig:sample_size_SnS}, we observe that shifts and their spread remain similar for all tested sample sizes, with only the smallest size (100 points) showing a noticeable deterioration in parameter estimation (larger shifts). On the other hand, the scatter of the posterior distributions, shown in the log--log plot in the bottom panel of Figure~\ref{fig:sample_size_SnS}, is clearly anticorrelated with the number of data points. The width of the posterior distribution scales with $N^{0.5}$, following the central limit theorem, where increasing $N$ reduces the variance of the posterior as $\sigma^2 / N$. This leads to a narrower $\pm1\sigma$ spread as more data points are included. To verify this, we included lines of the form $A/\sqrt{N}$, where $A$ was a scale factor obtained by fitting the line to the median values. These two panels in Figure~\ref{fig:sample_size_SnS} clearly demonstrate why using the scatter of posterior distributions as parameter uncertainties can be unreliable, which is a common practice in the literature. Even for light curves with 10{,}000 data points, the accuracy of parameter determination is similar to that for 500 points, while the posterior distribution can be up to an order of magnitude narrower.
\newline
\subsection{Dilution}
\label{sec:blending}
\begin{figure}
	\centering
	\includegraphics[width=1.0\linewidth]{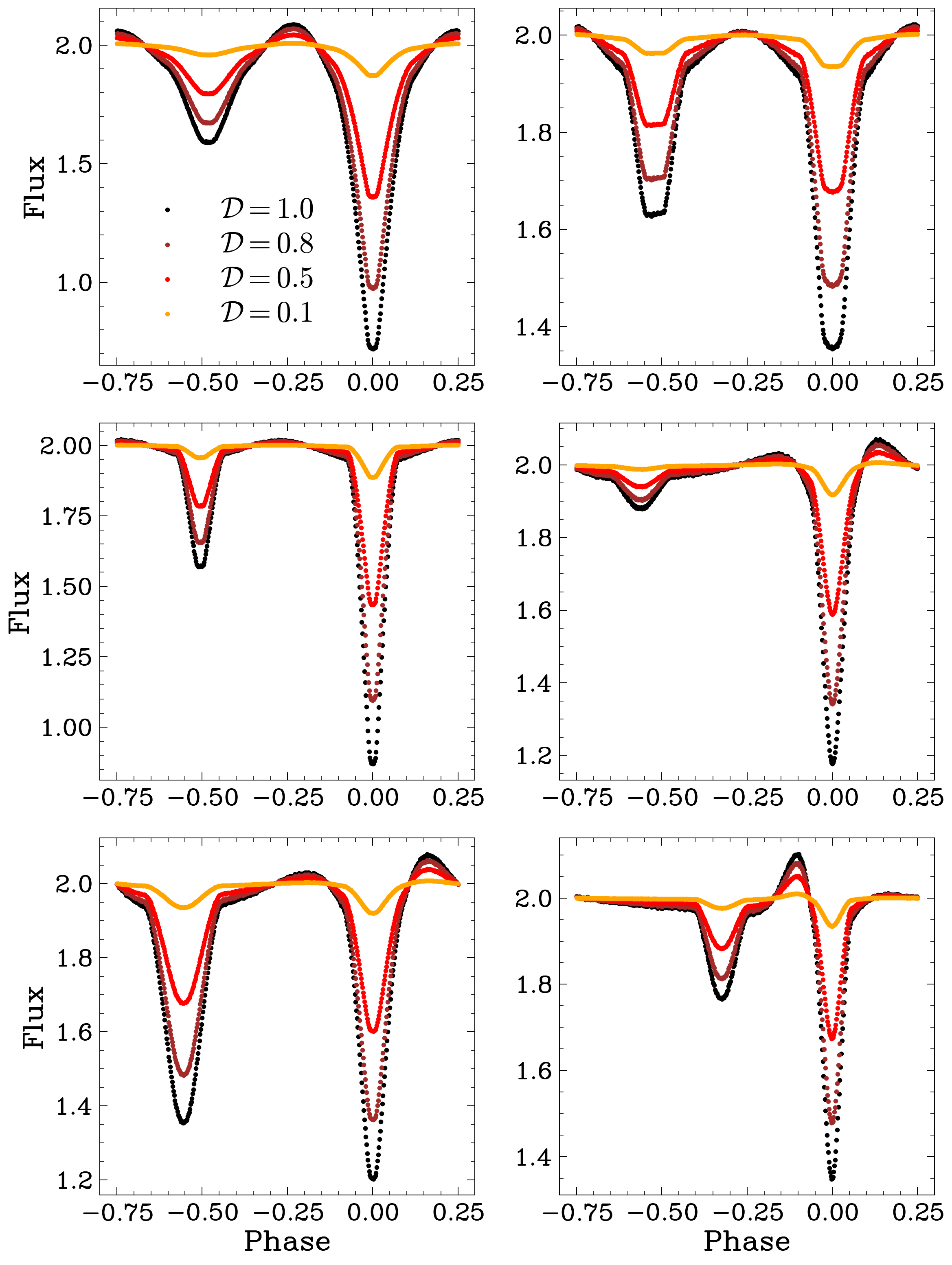}
	\caption{Synthetic light curves illustrating the effect of varying dilution fractions (\(\mathcal{D}\)) on light curve amplitudes.
	}
	\label{fig:blendinglcs}
\end{figure}

In this work, we treat the various effects of light contamination under a single term—dilution. This includes both additional light from sources that cannot be easily resolved from the system (e.g., a third body) as well as detector-dependent contamination, where the profile of a source star is diluted by a different source, for example, due to close proximity in the sky or the large pixel scale of the instrument. In the analysis of EBs, dilution mainly impacts the amplitude of variability because the additional flux increases the baseline, while the absolute change in flux remains the same. To better visualize the effect of dilution, we present several example light curves with various levels of dilution in Figure~\ref{fig:blendinglcs}. To describe the strength of the dilution effect, we use the dilution fraction ($\mathcal{D}$), which describes the fraction of total flux belonging to the system:
\begin{equation*}
	\mathcal{D} = \frac{F_{\rm S}}{F_{\rm S} + F_{\rm B}},
\end{equation*}
where $F_{\rm S}$ is the source flux and $F_{\rm B}$ is the additional flux due to the dilution effect. For $\mathcal{D} = 1.0$, there is no dilution (all flux comes from the system), while, e.g., $\mathcal{D} = 0.5$ means only half of the observed flux comes from the system.

To evaluate how dilution affects the accuracy of parameter determination, we ran two tests on the same sample of about 6000 newly generated light curves using the same setup as in previous tests. In the first test, we performed the standard modeling procedure. In the second test, we added $\mathcal{D}$ as a free parameter in the fitting process. For each object in our sample, we modeled 10 versions of the light curve with different dilution fractions, chosen from the set $(0.1, 1.0)$ in steps of 0.1.

\begin{figure}
	\centering
	\includegraphics[width=1.0\linewidth]{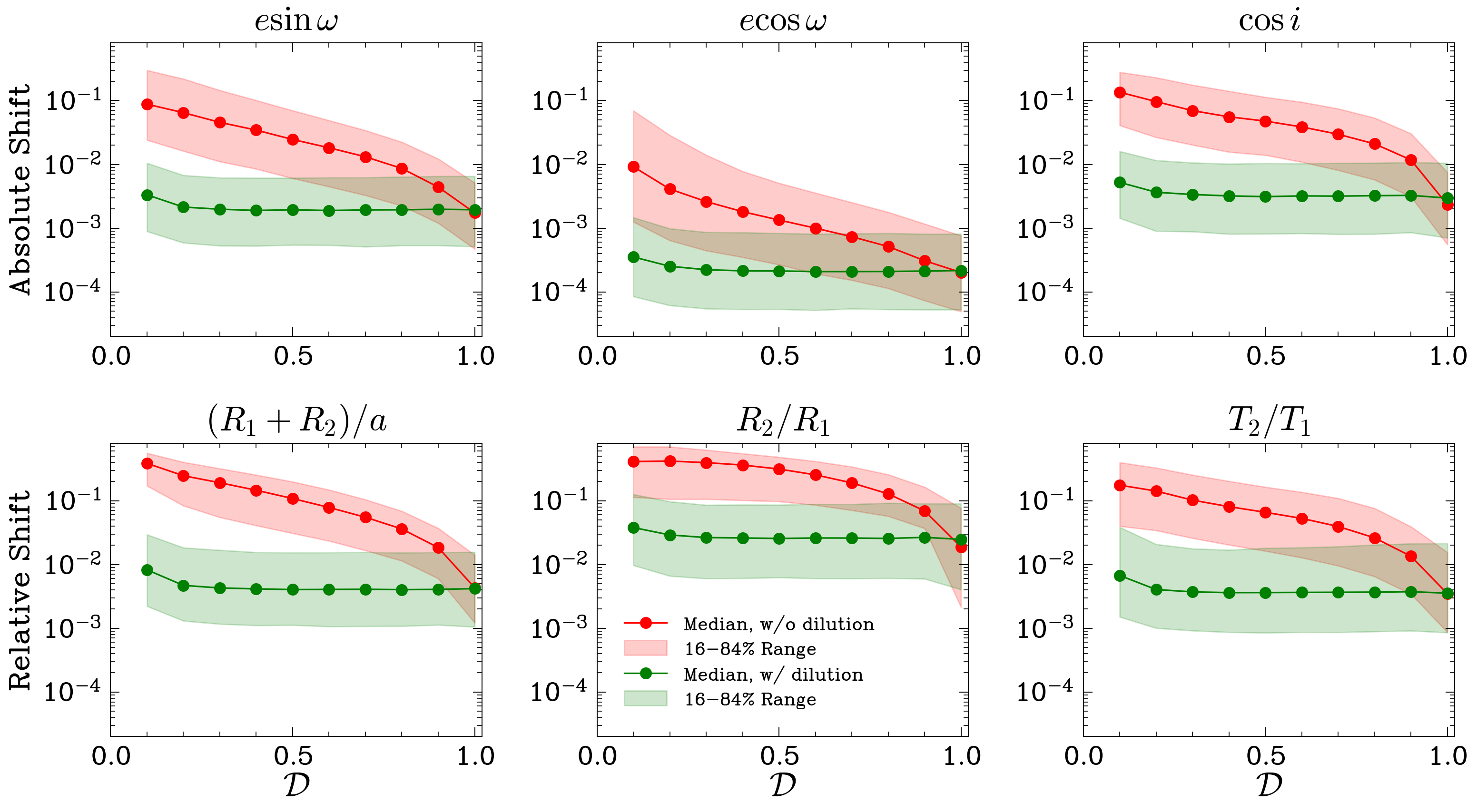}
	\caption{Median parameter shifts as a function of the dilution fraction. Red/green lines represent cases without/with \(\mathcal{D}\) as a free parameter in the model. For each \(\mathcal{D}\), we calculated the median and 16th–84th percentile range of the shift based on a sample of about 6000 synthetic light curves.
	}
	\label{fig:blending_shifts}
\end{figure}

Figure~\ref{fig:blending_shifts} shows the results of both tests. For each $\mathcal{D}$, we calculated the median and the 16th–84th percentiles of the distribution of shifts between the obtained parameter values and their true values. Including the dilution fraction significantly improves the accuracy of parameter determination. The first test demonstrates that, even with weak dilution, the obtained parameter values deviate significantly from the assumed values. The most striking example is $R_2/R_1$, which deteriorates rapidly even with minimal dilution. Parameter $e\cos\omega$ provides good estimates even under strong dilution because it affects the phase separation of the eclipses rather than the amplitude of brightness variations.

\begin{figure*}
	\centering
	\includegraphics[width=0.8\linewidth]{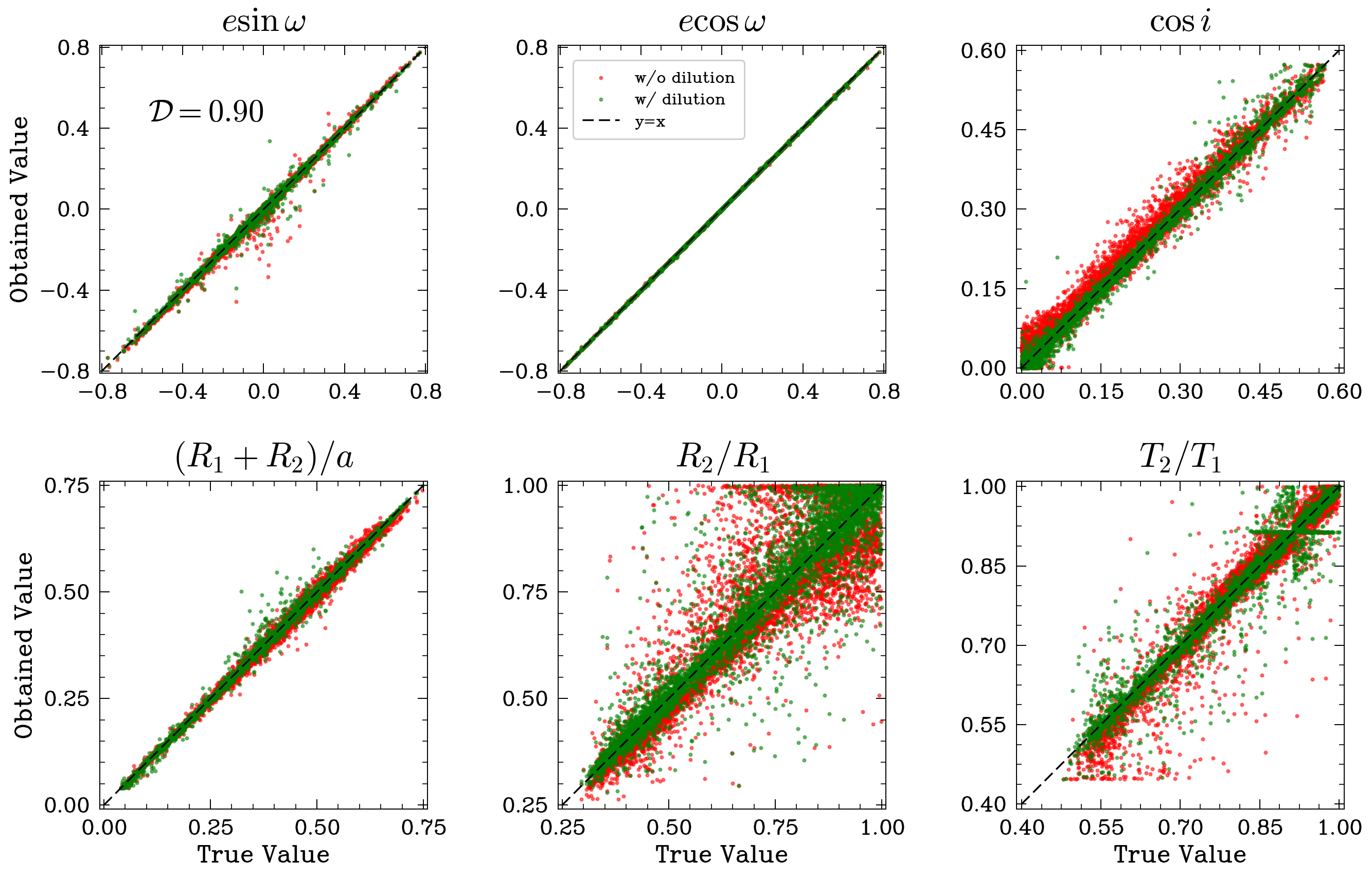}
	\textcolor{white}{\rule[0.5ex]{\linewidth}{0.5pt}}
	\includegraphics[width=0.8\linewidth]{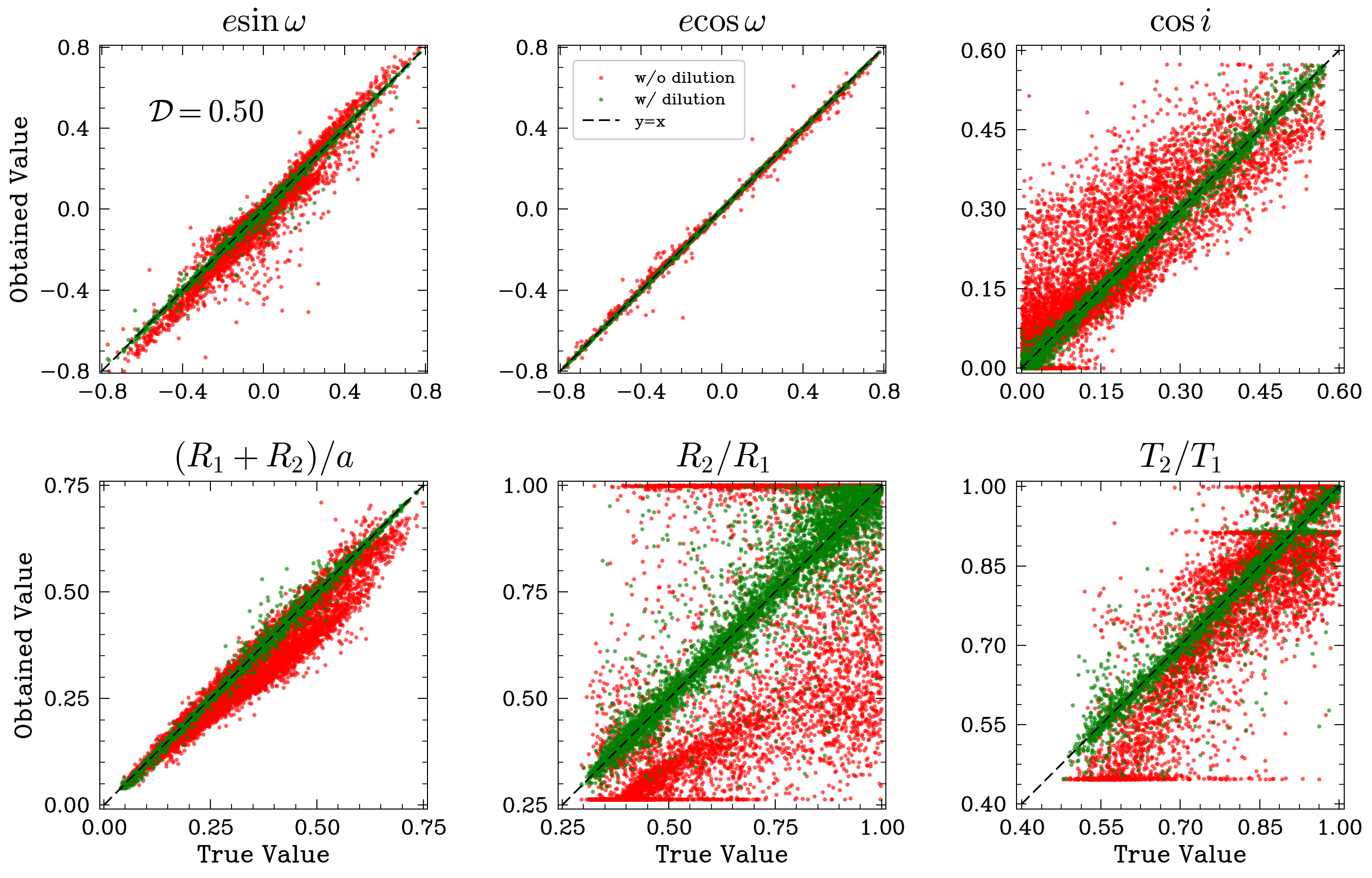}
	\caption{Comparison of true vs. recovered parameters after modeling about 6000 synthetic light curves using PHOEBAI. Green/red points represent results with/without the dilution factor as a free parameter. Black dashed lines show the \(y = x\) line. The upper/lower sets of plots show results for \(\mathcal{D} = 0.9\) / \(\mathcal{D} = 0.5\).
	}
	\label{fig:blendingtvo}
\end{figure*}

\begin{figure}
	\centering
	\includegraphics[width=1.0\linewidth]{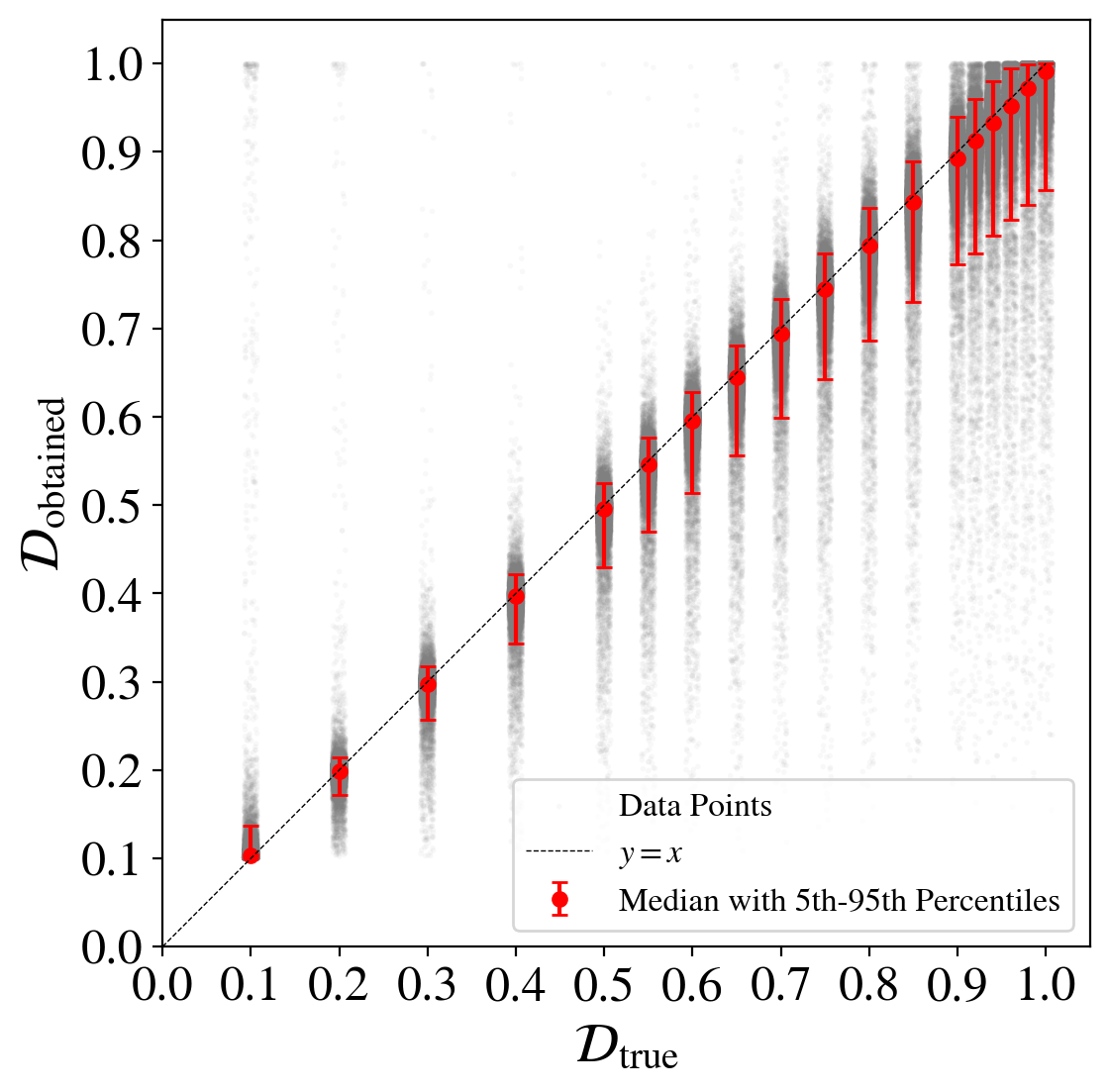}
	\caption{Comparison of true vs. predicted values of the dilution factor based on modeling 10{,}000 synthetic light curves. Red points denote the median value of the obtained \(\mathcal{D}\) for a given true value, while error bars represent the 5th and 95th percentiles of this distribution. To enhance visualization, a small horizontal random offset was applied to reduce overlap.}
	\label{fig:blendingdilution}
\end{figure}

For $\mathcal{D} = 1.0$, both tests returned consistent values. However, as $\mathcal{D}$ decreases, the differences between the two tests increase. Without $\mathcal{D}$ as a free parameter, changes in amplitude must be compensated by adjusting other parameters. Since the flux scaling factor, used in both tests, does not affect the relative amplitude of brightness variations, dilution-induced amplitude changes cannot be adjusted by modifying a single parameter, such as $\cos i$. While dilution affects the amplitude, it does not alter the ratio of primary to secondary eclipse amplitudes, requiring adjustments across several parameters. In contrast, treating $\mathcal{D}$ as a free parameter allows the program to accurately reproduce the assumed value of this parameter (see Fig.~\ref{fig:blendingdilution}) and recover other parameters with similar precision, regardless of the dilution level.

Figure~\ref{fig:blendingtvo} presents a comparison of true vs. recovered parameter values for two different dilution factors, $\mathcal{D} = 0.9$ (top) and $\mathcal{D} = 0.5$ (bottom). In both sets of plots, the second test (with $\mathcal{D}$ as a free parameter) produces very similar results, while the first test (without $\mathcal{D}$) shows drastic differences. Notably, for $\mathcal{D}=0.9$, the $R_2/R_1$ value saturates, with the program returning a boundary value. For $\mathcal{D}=0.5$, the lower limit is returned for nearly all objects. Other parameters, except $e\sin\omega$ and $e\cos\omega$, also exhibit significant loss of accuracy.

\begin{figure}
	\centering
	\includegraphics[width=1.0\linewidth]{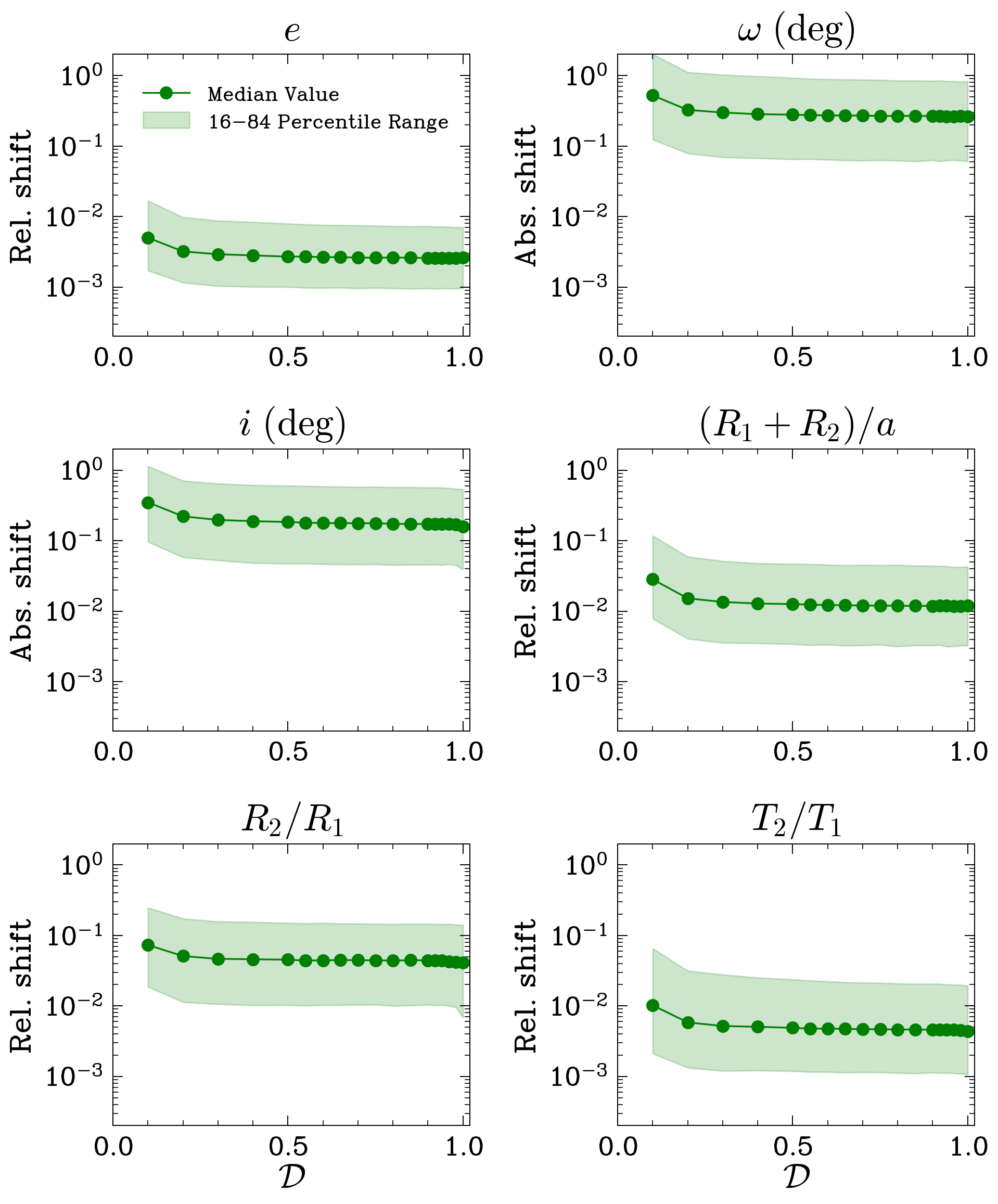}
	\caption{Shifts in the obtained parameters (\(e\), \(\omega\), \(i\) regime) relative to the dilution factor. All points represent modeling processes that included \(\mathcal{D}\) as a free parameter. For each \(\mathcal{D}\), we calculated the median and 16th–84th percentile range of shifts based on a sample of 10{,}000 synthetic light curves.
	}
	\label{fig:blending_shifts_v2_ewi}
\end{figure}

Introducing a new free parameter always risks the emergence of new degeneracies. To explore this, we extended the sample of synthetic light curves to approximately 10{,}000 and added new $\mathcal{D}$ values to the previous set. Figure~\ref{fig:blending_shifts_v2_ewi} presents the shifts for all parameters, including $e$, $\omega$, and $i$. A slight deterioration in accuracy is observed at both extremes—first, when comparing the results for light curves without dilution to those with even slight dilution, and second, under significant dilution (when only 10\% of the flux comes from the source). For intermediate values, the accuracy remains roughly constant.

\begin{figure}
	\centering
	\includegraphics[width=0.9\linewidth]{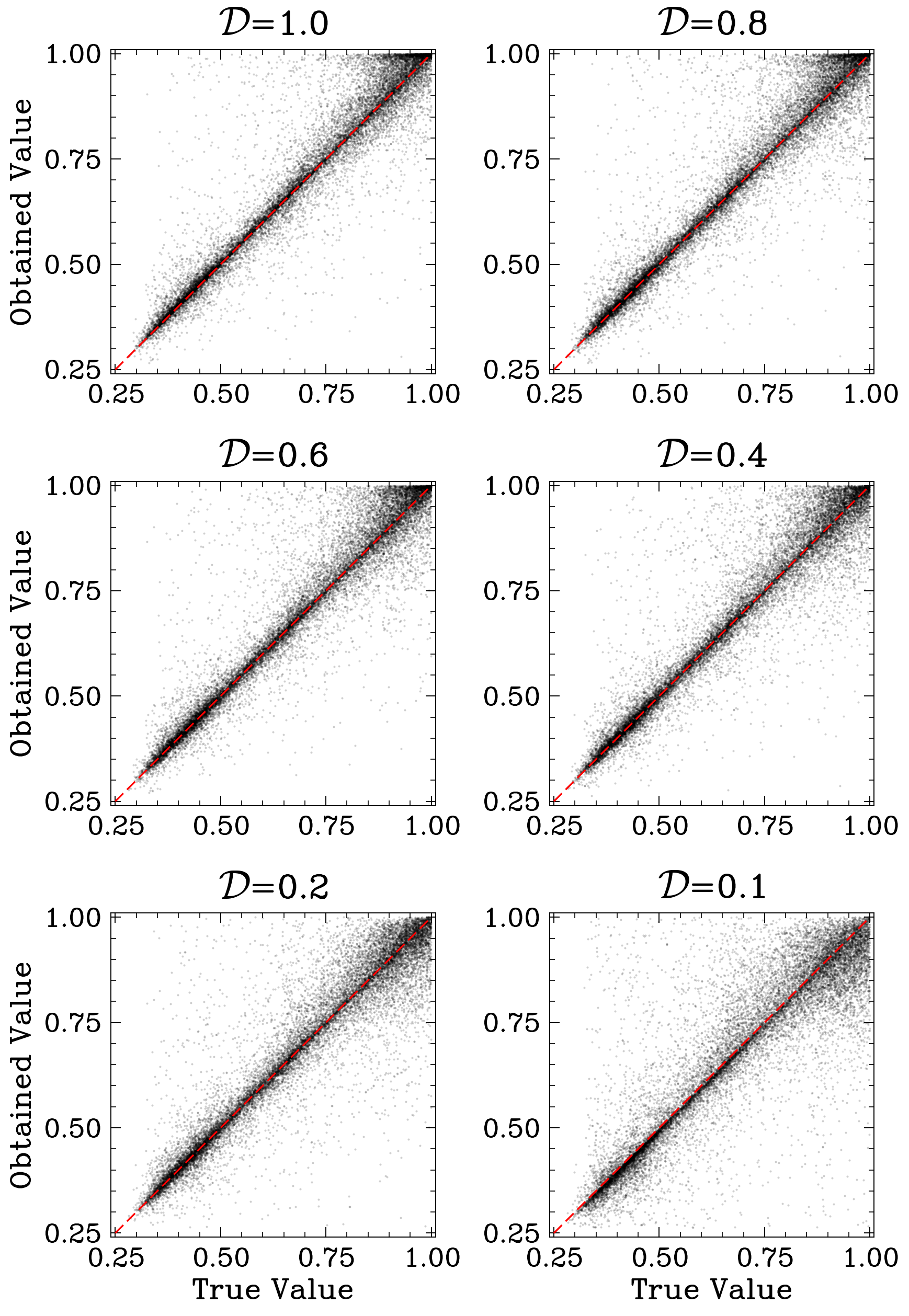}
	\caption{True vs. obtained values for the \(R_2/R_1\) parameter in relation to \(\mathcal{D}\).
	}
	\label{fig:blendingtvosingle}
\end{figure}

Figure~\ref{fig:blendingtvosingle} illustrates how increasing dilution levels influence the true vs. calculated plot for $R_2/R_1$. Similar plots for other parameters are not shown, as the changes across different $\mathcal{D}$ values are less pronounced. We observe that for all cases, except the one with the lowest dilution fraction, the plots remain similar. Figure~\ref{fig:blendingdilution} depicts the true vs. restored values for $\mathcal{D}$. Most points align along the identity line, but some cases show significant deviations. Particularly curious are cases where no dilution effect is present, but the program returns solutions with a nonunity dilution factor. Such degeneracies are unfortunately difficult to avoid when introducing an additional free parameter into the model. A simple solution to prevent the incorrect determination of the model parameters is to avoid using the dilution fraction as a free parameter if there is strong confidence that a single source is being observed. Another solution is to impose a prior distribution on $\mathcal{D}$, which would depend on the density of the observed field and the brightness of the examined object.

As demonstrated, in most cases, accounting for the dilution effect in the analysis is essential for retrieving accurate parameter values. Even with slight light contamination, the determined parameters can exhibit large errors. This effect is particularly significant when modeling the light curves of EBs from photometric surveys, which are often prone to dilution. This is especially relevant for projects using cameras with large pixel scales, such as TESS (21''\,px$^{-1}$) or Kepler/K2 (3.98''\,px$^{-1}$), or for surveys observing densely populated regions of the sky, like the OGLE project monitoring the Galactic bulge.

\section{Discussion}
The results from our tests are highly promising, and with the validation that the ANNs can reliably stand in for the complex physical model, we are now prepared to apply this methodology to actual systems. While the ANN architecture and training dataset size can remain unchanged, we will need to train multiple networks to cover the large range of parameter combinations, such as the primary component's effective temperature, radius, and mass, as well as different passbands and the specific binary system configurations (e.g., two main-sequence stars versus systems where one component has evolved to the red-giant branch).

Our analysis has established a threshold for the practical accuracy of the applied ANN. A useful analogy here is to imagine the model produced by the ANN as having an intrinsic noise level. This is clearly illustrated in the top panel of Figure~\ref{fig:sigma_vs_shift_scatter}—beyond a certain level of injected noise, the accuracy of derived results does not improve, consistently across each network. For the main parameter set, this threshold level of noise can be approximated as around $10^{-3}$, representing the ANN’s “intrinsic noise.” The achieved accuracy, which can also be regarded as a systematic error, varies by specific parameter. The parameter $\ecosw$ can be determined with absolute accuracy down to $2\times10^{-5}$, whereas $(R_1 + R_2)/a$ exhibits a less accurate relative accuracy, typically no better than approximately 0.1\% even under ideal conditions. Table~\ref{table:precision_statistics} provides statistics for each parameter, where for each of the five ANNs, we calculated the mean accuracy across all data points with a noise level below or equal to $10^{-3}$, reporting the minimum, median, and maximum accuracy across all ANNs.

\begin{table}[h!]
	\centering
	\caption{Precision statistics for each parameter across all ANNs, with noise levels below or equal to $10^{-3}$. The table reports minimum, median, and maximum accuracy for each parameter.}
	\begin{tabular}{lccc}
		\hline
		Parameter & Min & Median & Max \\
		\hline
		$\esinw$ & 1.03e-03 & 2.36e-03 & 3.27e-03 \\
		$\ecosw$ & 8.04e-05 & 9.65e-05 & 3.95e-04 \\
		$\cosi$ & 2.84e-04 & 7.33e-04 & 1.94e-03 \\
		$(R_1+R_2)/a$ & 1.10e-03 & 3.03e-03 & 9.12e-03 \\
		$\rratio$ & 5.39e-04 & 1.08e-03 & 5.61e-03 \\
		$\tratio$ & 6.29e-04 & 2.03e-03 & 3.56e-03 \\
		\hline
	\end{tabular}
	\label{table:precision_statistics}
\end{table}

Based on Figures~\ref{fig:sigma_vs_shift_scatter} and~\ref{fig:sample_size_SnS}, estimating the measurement uncertainties of derived parameters solely from the posterior distribution spread can be misleading. Even a well-sampled light curve with minimal uncertainty per data point does not allow parameter estimation beyond the systematic error determined earlier, despite potentially low standard deviations in the posterior distribution from the MCMC sampling suggesting otherwise. Conversely, when the available data points are limited (fewer than 1000) or when the light curve exhibits substantial noise (with relative levels exceeding 0.5~\(\times 10^{-4}\)), statistical error begins to dominate over the systematic error, which is apparent in the declining parameter accuracy in the upper panels of both figures mentioned above.

One of the main motivations for this study was to improve the speed of obtaining the estimation of the main set of parameters and their posterior PDFs. To compare the times in the most comprehensive way, we used the exact settings for both samplers (the same number of walkers and epochs, which were 80 and 500, respectively, and the same set of initial vectors) and the same hardware configuration (96 CPUs, Intel(R) Xeon(R) Gold 6240R CPU @ 2.40GHz). The average times for both samplers were \(2.16 \pm 0.36\,\mathrm{s}\) for ANNs and \(21.63 \pm 1.41\,\mathrm{hr}\) for PHOEBE. Thus, at a given stage, the speedup is above \(10^4\) times.

\section{Conclusions}
To address the pressing issue of high computational demands for fully modeling the light curves of eclipsing binary (EB) systems, we developed an ANN capable of efficiently predicting forward models based on a defined set of principal parameters, specifically \( e\sin\omega \), \( e\cos\omega \), \( \cos i \), \( (R_1 + R_2)/a \), \( R_2/R_1 \), and \( T_2/T_1 \).

Our initial objective was to identify an ANN architecture that achieves an optimal balance between accuracy and computational efficiency. Utilizing the \texttt{RandomizedSearchCV} method, we explored various ANN configurations to determine how well they predicted forward models, in this case, a regression task focused on predicting flux values based on input parameters. We found that the ideal architecture comprises six hidden layers, each with 512 nodes. The activation function for the first hidden layer is \texttt{elu}, with \texttt{sigmoid} functions for the remaining hidden layers and a \texttt{linear} activation function at the output layer. This configuration proved effective in capturing the complexity of the problem while maintaining high efficiency, and we identified potential optimization patterns applicable to similar regression tasks (see Section~\ref{sec:structure_of_DNN}).

Prior to applying the ANN to real EB data, we verified its accuracy in modeling synthetic light curves. This controlled setup, where all physical and orbital parameters were known, allowed us to identify and quantify systematic errors, which might otherwise have gone undetected. Using a large dataset of 1.25 million generated models, we conducted cross-validation tests to confirm that the ANN’s performance is not significantly dependent on the specific training subset used (see Section~\ref{sec:CV_check}). For a randomly selected synthetic light curve sample, we compared the results from the ANN to those obtained through traditional modeling with PHOEBE, demonstrating strong agreement.

In further analyses, we assessed the impact of data uncertainty (Section~\ref{sec:data_uncertain}) and sample size (Section~\ref{sec:data_size}) on the accuracy of parameter estimates and the scatter in their posterior distributions. We found that the systematic error is comparable to white noise, with a relative sigma level of approximately one part per thousand. Additionally, the average posterior distribution scatter scales with the noise level. We observed that, beyond a sample size of 200 data points, the number of points does not significantly affect parameter accuracy. However, the posterior scatter is inversely proportional to the square root of the sample size, in alignment with theoretical expectations.

Using an unseen set of 100{,}000 generated models, we applied the ANN modeling process to investigate potential systematic errors related to parameter values within the primary set (Section~\ref{sec:params_value}). With few exceptions, the accuracy of derived parameters generally does not show notable dependency on the parameter values themselves. The most significant deviation was observed for the \( R_2/R_1 \) parameter when comparing systems with and without total eclipses; in the latter case, the relative accuracy improved by up to an order of magnitude. We also noted reduced relative accuracy for \( e \) and \( (R_1 + R_2)/a \), likely due to their lower limit proximity to zero.

Additionally, we explored the impact of light contamination (dilution effect) on parameter accuracy (Section~\ref{sec:blending}). Even minimal dilution caused the ANN-derived results to deviate substantially from the expected values. We quantified dilution through a dilution fraction, \( \mathcal{D} \), and addressed the issue by including \( \mathcal{D} \) as an additional fitting parameter, applied post-ANN light curve prediction. While introducing an extra parameter could risk new degeneracies, we found that, for the majority of tested synthetic light curves, the model effectively recovered both the true values of the main parameters and the correct \( \mathcal{D} \) value, even under strong dilution conditions.

Our tests demonstrated that the current ANN implementation achieves a speedup of over four orders of magnitude. However, there remains potential for further optimization through refining the ANN architecture and employing more advanced algorithms for forward-model computations, which could enhance this speedup even further. While this acceleration primarily benefits the sampling process, it is important to note that the estimation and optimization of model parameters also gain from bypassing forward-model computations with PHOEBE.

A natural next step is to apply this optimized ANN framework to estimate the primary parameter set for detached EBs cataloged in surveys such as OGLE, TESS, and Kepler/K2. Achieving this, however, will require training new networks on datasets generated over a broader parameter range, particularly for parameters with secondary effects on light curve morphology. Additionally, successful parameter estimation will necessitate manual inspection and model adjustments to account for uncommon or exotic EBs, whether using a physical model directly or an ANN as a stand-in. 

\section*{Acknowledgments}
The development of PHOEBE has been made possible through the NSF AAG grants \#1517474 and \#1909109 and NASA 17-ADAP17-68, which we gratefully acknowledge. This research was funded by NASA grant 23-ADAP23-0068 and carried out as part of the research project No. BPN/BEK/2023/1/00067, funded by the Polish National Agency for Academic Exchange under the Bekker NAWA 2023 program.

\vspace{5mm}

\software{Astropy \citep{astropy:2018}, 
        matplotlib \citep{2007CSE.....9...90H},
        numpy   \citep{numpy},
        scipy \citep{2020SciPy-NMeth},
}

\bibliography{bibfile}{}
\bibliographystyle{aasjournal}

\end{document}